\documentclass[a4paper, 12pt]{article}

\usepackage{epsfig, psfrag}

\usepackage{afterpage}

\usepackage{amsfonts, amsmath, amssymb, amsthm}

\usepackage{tabls}
\usepackage{multirow}
\usepackage{rotating}

\usepackage{bbm}

\newcounter{thmNumber}[section]
\newcounter{lemmaNumber}[section]
\newcounter{corollaryNumber}[section]

\numberwithin{thmNumber}{section}
\numberwithin{lemmaNumber}{section}
\numberwithin{corollaryNumber}{section}
\numberwithin{definitionNumber}{section}
\newtheorem{theorem}[thmNumber]{Theorem}

\newtheorem{corollary}[corollaryNumber]{Corollary}
\newtheorem{lemma}[lemmaNumber]{Lemma}

\newcommand{\R}{{\mathbb R}}
\newcommand{\C}{{\mathbb C}}
\newcommand{\Z}{{\mathbb Z}}
\newcommand{\N}{{\mathbb N}}

\newcommand{\be}{\begin{equation}}
\newcommand{\ee}{\end{equation}}

\newcommand{\Rhat}{\hat{R}}
\newcommand{\LL}{{\bf\underline{L}}}
\newcommand{\SSS}{{\bf\underline{S}}}
\newcommand{\Etilde}{\tilde{E}}
\newcommand{\Span}{\,\mbox{Span}\,}

\begin{document}

\title{Explicit large nuclear charge limit of electronic ground states for Li, Be, B, C, N, O, F, Ne
and basic aspects of the periodic table}

\author{Gero Friesecke
        and Benjamin D. Goddard}

\date{January 16, 2009}

\maketitle

\begin{abstract} This paper is concerned with the Schr\"odinger equation for atoms and ions with
$N=1$ to 10 electrons. In the asymptotic limit of large nuclear charge $Z$,
we determine explicitly the low-lying energy levels and eigenstates.

The asymptotic energies and wavefunctions are in good quantitative agreement with experimental data for positive ions,
and in excellent qualitative agreement even for neutral atoms ($Z=N$). In particular, the predicted ground state spin and angular momentum
quantum numbers ($^1S$ for He, Be, Ne, $^2S$ for H and Li, $^4S$
for N, $^2P$ for B and F, and $^3P$ for C and O) agree with experiment in every case.

The asymptotic Schr\"odinger ground states agree, up to small corrections, with the semi-empirical
hydrogen orbital configurations developed by Bohr, Hund and Slater to explain the periodic table.
In rare cases where our results deviate from this picture, such as the ordering of
the lowest ${}^1D^o$ and ${}^3S^o$ states of the Carbon isoelectronic
sequence, experiment confirms our, not Hund's, predictions.

\end{abstract}

\section{Introduction}
How do the striking chemical differences between some elements, and the similarities between others,
emerge from the universal laws of quantum mechanics? In the physics and chemistry literature,
this fundamental question is discussed on a semi-empirical level, via the ``hydrogen orbital configurations''
developed by Bohr, Hund and Slater (see e.g. \cite{Bohr22, Hund25, Slater30, CondonShortley35, LanLif, Schwabl01, Atkins01}), or via numerical simulation
of simplified quantum mechanical models (see e.g. \cite{Hartree28, Hartree57, FroeseFischer77, Tatewaki94, NIST05, Tayloretal1}).
In this article, we address this question from a mathematical perspective.

The theoretical possibility of making chemically specific predictions
was realized almost immediately after
the Schr\"odinger equation had been introduced (see e.g. \cite{Hartree28, Dirac29}).
But we are not aware of previous rigorous results, as mathematical research
on the basic quantum mechanical equations has hitherto focused overwhelmingly
on universal, qualitative properties.

The perhaps most basic non-universal properties of atoms relevant to
chemical behaviour are the total spin and angular momentum quantum
numbers $S$ and $L$ of the ground state, which describe the amount
of symmetry under spin and spatial rotation. These two numbers not
only determine the ground state dimension $d$, but, as argued below,
they also allow to predict, up to at most two possibilities, the
group of the atom in the periodic table.

Other quantities of interest include the energy levels $E_n$ and, more importantly,
energy differences such as spectral gaps
$E_n-E_m$ (which govern the photon frequencies which the atom can emit or absorb)
and the ionization energy (whose striking empirical periodicities
lay at the origin of the design of the periodic table).

Our principal result is that such quantitites can be extracted analytically
from the many-electron Schr\"odinger equation in a natural scaling limit.
More precisely, we show that for ions with $N=1$ to 10 electrons,
as the nuclear charge $Z$ gets large the low-lying energy levels
and eigenstates converge to well defined limits,
which can be determined explicitly.
In particular, this yields rigorous values of $L$, $S$, $d$ for the ground state for all sufficiently large $Z$.
See Theorems \ref{GStheorem}, \ref{T:isolimit}, \ref{mainresult} and Tables \ref{Tab:GSqnumbers}, \ref{Tab:PTGS}, \ref{Tab:PTEigenspaces1},
\ref{Tab:PTEigenspaces2}. We call the above fixed-$N$, large-$Z$ limit {\it iso-electronic limit}, because
it is realized physically by an iso-electronic sequence such as Li, Be$^+$, B$^{++}$, ... Note that this limit
is different from the Thomas-Fermi limit $N=Z\to\infty$, which is
of interest in other contexts but does not retain any chemical specificity.

The asymptotic levels and eigenstates are in good quantitative agreement with experimental data for positive ions,
and in excellent qualitative agreement even for neutral atoms ($Z=N$). In particular
the predicted values of $L$, $S$ and $d$ (see Table \ref{Tab:GSqnumbers}) agree with the experimental atomic values
\cite{NIST} in all cases.

\begin{table}[ht]
 \begin{center}
 \resizebox{0.8\textwidth}{!}{
  \begin{tabular}{|c|c|c|c|c|c|c|c|c|c|c|}
   \hline
   {Iso-electronic sequence} & H & He& Li&Be  & B  & C  & N  & O  & F & Ne \\
   \hline
   $\sharp$ electrons & 1 & 2 & 3 & 4  & 5  & 6  & 7  & 8  & 9 & 10 \\
   \hline\hline
   $L$ & 0 & 0 & 0 & 0 & 1 & 1 & 0 & 1 & 1 & 0 \\
   \hline
   $S$ & $\frac{1}{2}$ & 0 & $\frac{1}{2}$ & 0 & $\frac{1}{2}$ & 1 & $\frac{3}{2}$ & 1 & $\frac{1}{2}$ & 0 \\
   \hline
   {\tiny Chemist's notation} & {\tiny $^2S$} & {\tiny $^1S$} & \tiny{$^2S$} & {\tiny $^1S$} & {\tiny $^2P$} &
    {\tiny $^3P$} & {\tiny $^4S$} & {\tiny $^3P$} & {\tiny $^2P$} & {\tiny $^1S$} \\
   \hline
   dim & 2 & 1 & 2 & 1 & 6 & 9 & 4 & 9 & 6 & 1 \\
   \hline
  \end{tabular}
  } 
 \end{center}
\caption{Angular momentum and spin quantum numbers and dimension of the Schr\"odinger ground state for large $Z$,
  as calculated in this paper.
  All numbers agree with the experimental values for neutral atoms \cite{NIST}.}
 \label{Tab:GSqnumbers}
\end{table}

The asymptotic ground states we calculate (see Theorem \ref{GStheorem}) provide for the first time a mathematical
justification of the celebrated semi-empirical ``hydrogen orbital configurations'' developed
notably by Bohr, Hund and Slater to explain the periodic table. In our approach,
none of the underlying nontrivial postulates (electrons filling hydrogen
orbitals, shell and sub-shell formation, sub-shell ordering rules
such as 2s$<$2p, Hund's rules) need to be invoked, but are seen to emerge in a natural way.
The only corrections are as follows (see Sections \ref{Sec:Aufbau} and \ref{Sec:Comparison} for a detailed discussion):

(1) Alongside each Slater determinant built from
admissible hydrogen orbitals, the ground state must contain its orbit under the symmetry group $SO(3)\times SU(2)\times \Z_2$
of the many-electron Schr\"odinger equation.

(2) For the three elements Be, B, C, a ten to twenty percent admixture of a particular ``higher sub-shell''
configuration is also present, an effect we term $2s^2$--$2p^2$ resonance.

(3) In rare cases, such as that of the
lowest ${}^1D^o$ and ${}^5S^o$ states of Carbon, the ordering of excited states disagrees with
the semi-empirical Hund's rules (with experiment confirming our orderings).

We now outline our mathematical strategy to obtain explicit asymptotic energy levels and eigenstates, focusing for
simplicity on the ground state.

The first step is the derivation of a simplified model governing the asymptotics. This can be done via
a scaling argument plus standard perturbation theory, as follows.
For fixed $N$ and large $Z$, attraction of an electron by the nucleus dominates
its interaction with the other electrons, so one expects
the true ground state to be close to the ground state of the corresponding system with electron interaction turned off
(which is known explicitly via hydrogen atom theory). After a little more thought, one realizes that this cannot be quite correct.
The non-interacting ground state eigenspaces of atoms
happen to be highly degenerate (see Table \ref{Tab:H0Dim}), but the underlying symmetry is broken by
the interaction, so the true ground state eigenspaces should
converge only to particular subspaces of them.
(Experimentally, this phenomenon is well known, from observed energy splittings.)
Mathematically, we prove that the difference between the Schr\"odinger ground states and the ground states
of the problem $PHP\Psi=E\Psi$, where $P$ is the projector onto the non-interacting ground state but
$H$ is the full Hamiltonian (eq. (\ref{ham}) below), tend to zero. We call this simplified problem {\it PT model},
because it corresponds to (i) re-scaling the problem so as to make the ground state of the reduced problem
$Z$-independent, (ii) applying degenerate first order perturbation theory, (iii) undoing the re-scaling.
Physically it corresponds to resolving, within the non-interacting ground state eigenspace,
the fully interacting problem.

The second step is to determine the lowest PT eigenspace. This requires a careful analysis of the interplay
between hydrogen orbital formation (promoted by the Laplacian and electron-nuclei interaction), antisymmetry,
spin, and electron interaction. More technically,
the following difficulties arise. \\
(i) The non-interacting ground state, i.e. the state space of the PT model, is of somewhat daunting looking
dimension, e.g. 70 in case of Carbon (see Table \ref{Tab:H0Dim}). \\
(ii) The PT Hamiltonian $PHP$ is easy to write down abstractly (as we have just done), but unknown; one needs to
devise a method to determine it explicitly. \\
(iii) The PT model is a strongly interacting many-body model. \\
Difficulties (i) and (iii) are overcome via careful use of the
symmetry group of the original equation and its representation theory in terms of many-body spin and
angular momentum operators, which allows one to split the Hamiltonian $PHP$ into small invariant blocks. (ii) is addressed by combining
ideas from quantum chemistry which have not hitherto played a role in mathematical studies, such as Slater's rules \cite{SzaboOstlund96}
(which allow to express the components of the Hamiltonian via six-dimensional integrals of a product of four hydrogen eigenstates
and a Coulomb repulsion term), Fourier analysis (while the Fourier transform of individual hydrogen eigenstates is well known,
here one requires the Fourier transform of pointwise products of these), and residue calculus. In principle, our methods apply to
arbitrary atoms, except that the relevant $PHP$ matrices can become significantly higher dimensional.

One curious mathematical phenomenon we observe is that the Hamiltonian
$PHP$ arising in the $Z\to\infty$ limit of quantum mechanics is always a rational matrix,
despite $H$ being a somewhat complicated partial differential operator and
$P$ a ``transcendental'' projector (onto tensor products of scaled
hydrogen eigenfunctions such as $\pi^{-1/2}e^{-|x|}$).

The structure of this paper is as follows. In Section 2 we describe the basic quantum mechanical equations and their
symmetry group. In Section 3 we state the asymptotic limit of the Schr\"odinger ground states for
Li to Ne (see Theorem \ref{GStheorem}). In Sections 4 and 5, we justify the reduction to the
PT model and determine explicitly its state space.
Sections 6--7 contain more technical material: the explicit determination of the PT Hamiltonian and
the derivation of Theorem \ref{GStheorem}, as well as of the excited states and levels of the PT model. Finally,
sections \ref{Sec:Aufbau} and \ref{Sec:Comparison} compare
our results to experimental data and to methods in the physics and chemistry literature.

\section{Schr\"odinger equation and mathematical definition of basic quantitites of chemical physics} \label{Sec:Basic}
The exact (nonrelativistic, Born-Oppenheimer) time-independent Schr\"odinger equation
for atoms and ions is
\be \label{SE}
    H \Psi = E \Psi,
\ee
where, for nuclear charge $Z>0$ and $N$ electrons and in atomic units,
\be \label{ham}
    H = \sum_{i=1}^N \Big( -\frac{1}{2} \Delta_{x_i} - \frac{Z}{|x_i|}
    \Big) + \sum_{1 \leq i < j \leq N} \frac{1}{|x_i-x_j|},
\ee
$E\in\R$, and
\be \label{space}
    \Psi \in L^2_a\big((\R^3 \times \Z_2)^N\big).
\ee
Here and below the $x_i\in\R^3$ are electronic coordinates, $s_i\in\Z_2=\{\pm\frac12\}$ are spin coordinates,
and $L^2_a$ is the usual Hilbert space of $N$-electron
functions $\Psi\, : \, (\R^3\times\Z_2)^N\to\C$ which are square-integrable,
\be \label{norm}
   \int_{\R^{3N}}\sum_{(\Z_2)^N} |\Psi(x_1,s_1,\dots,x_N,s_N)|^2 = ||\Psi||^2 <\infty,
\ee
and satisfy the antisymmetry principle that, for all $i$ and $j$,
\be \label{anti}
    \Psi(\dots, x_i,s_i, \dots, x_j,s_j, \dots)= -\Psi(\dots, x_j,s_j, \dots, x_i,s_i, \dots).
\ee
Mathematically, $H$ is a bounded below, self-adjoint operator with domain $L^2_a\cap H^2$, where
$H^2$ is the usual Sobolev space of $L^2$ functions with second weak derivatives belonging to $L^2$ \cite{Kat51}.

We are interested in the mathematical derivation of a number of quantities of basic physical and chemical interest,
and begin by recalling how these are defined in terms of the Schr\"odinger equation (\ref{SE}).
\\[2mm]
{\bf Definitions, 1} An {\it energy level of an atom or ion} is an eigenvalue of the corresponding operator $H$.
An {\it eigenstate of the atom or ion} is an eigenstate of $H$ (i.e. a nonzero solution $\Psi$ to (\ref{SE})
belonging to the domain of $H$). By Zhislin's theorem (\cite{Zhislin60}, see \cite{Friesecke03} for a short proof),
for atoms ($N=Z$) and positive ions ($N<Z$) there exist countably many energy levels $E_1<E_2<...$ below the bottom of
the essential spectrum of $H$, the corresponding eigenspaces being finite-dimensional. $E_1$ is called the {\it ground state
energy} and the corresponding eigenspace is known as the {\it ground state}. Eigenspaces corresponding to the higher energy levels
are known as {\it excited states}. The {\it excitation energy} or {\it spectral gap} of an excited state with energy $E_m$ is
defined to be $E_m-E_1$. Physically it corresponds to the energy required to promote the electrons from the ground state to the
excited state.
\\[2mm]
Besides the quantized energy levels $E_n$, there exist important additional discrete quantum numbers associated with
the atomic Schr\"odinger equation
which arise from its symmetries. Their precise definition, albeit very natural, takes
a little more work.
\\[2mm]
The model (\ref{SE}), (\ref{ham}), (\ref{norm}), (\ref{anti}) is invariant under
\begin{itemize}
\item [(i)] simultaneous rotation of all electron positions about the origin, \\
$\Psi(x_1,s_1,..,x_N,s_N)\mapsto\Psi(R^Tx_1,s_1,..,R^Tx_N,s_N)$, $R\in SO(3)$
\item [(ii)] simultaneous rotation of all electron spins \\
(by a unitary matrix $U\in SU(2)$)
\item [(iii)] simultaneous inversion of all electron positions at the origin, \\
$\Psi(x_1,s_1,..,x_N,s_N)\mapsto\Psi(-x_1,s_1,..,-x_N,s_N)=:\Rhat
\Psi$.
\end{itemize}
(In group theory language, the symmetry group is $SO(3)\times SU(2) \times \Z_2$, the third factor being the inversion group
consisting of $\Rhat$ and the identity. When $N=1$, there exists an additional symmetry, which gives rise to conservation of
the quantized Runge-Lenz vector; but it is broken by the interaction term in (\ref{ham}) when $N\ge 2$. Note also that even though
the Hamiltonian (\ref{ham}) is invariant under the larger group of non-simultaneous rotation of spins,
the antisymmetry condition (\ref{anti}) is not.)

The conserved quantities, i.e., operators which commute with the Hamiltonian, which arise from the above symmetries are
\begin{itemize}
\item[(i)]
   $\LL=\sum_{j=1}^N\LL(j)$ (many-electron angular momentum operator)
\item[(ii)]
   $\SSS=\sum_{j=1}^N\SSS(j)$ (many-electron spin operator)
\item[(iii)]
   $\Rhat$ (parity operator),
\end{itemize}
where
$$
    \LL(j)=\begin{pmatrix} L_1(j) \\ L_2(j) \\ L_3(j) \end{pmatrix},
    \quad
    \SSS(j)=\begin{pmatrix} S_1(j) \\ S_2(j) \\ S_3(j) \end{pmatrix},
$$
and $L_\alpha(j)$, $S_\alpha(j)$ ($\alpha=1,2,3$) denote the usual angular momentum respectively spin operators acting on the $j^{th}$
coordinate. Explicitly, on $N$-electron states $\Psi(x_1,s_1,..,x_N,s_N)$, $x_j\in\R^3$, $s_j\in\{\pm \frac12\}$, and denoting
$x_j=(y^{(1)}, y^{(2)}, y^{(3)})$, $L_\alpha(j)$ is the partial differential operator
\be \label{Lj}
   L_\alpha(j) = \frac1i \Bigl( y^{(\alpha+1)}\frac{\partial}{\partial y^{(\alpha-1)}} - y^{(\alpha-1)}\frac{\partial}{\partial y^{(\alpha+1)}}
   \Bigr),
\ee
and $S_\alpha(j)$ is multiplication by a Pauli matrix,
$$
   \left(\begin{array}{c} (S_\alpha(j)\Psi)(x_1,s_1,\ldots,
    x_j,\mbox{$\frac12$},\ldots,x_N,s_N) \\ (S_\alpha(j)\Psi)
    (x_1,s_1,\ldots,
    x_j,-\mbox{$\frac12$},\ldots,x_N,s_N)
   \end{array}\right)
   = \sigma_\alpha
   \left(\begin{array}{c} \Psi(x_1,s_1,\ldots,
    x_j,\mbox{$\frac12$},\ldots,x_N,s_N) \\ \Psi(x_1,s_1,\ldots,
    x_j,-\mbox{$\frac12$},\ldots,x_N,s_N)
   \end{array}\right),
$$
where the $\sigma_\alpha$ are the Pauli matrices
$$
    \sigma_1:=\frac12\begin{pmatrix} 0 & 1 \\ 1 & 0 \end{pmatrix}, \quad
    \sigma_2:=\frac12\begin{pmatrix} 0 & -i \\ i & 0 \end{pmatrix}, \quad
    \sigma_3:=\frac12\begin{pmatrix} 1 & 0 \\ 0 & -1 \end{pmatrix}.
$$
The fact that the operators (i), (ii), (iii) commute with the Hamiltonian (\ref{ham})
can be checked by direct inspection using the above formulae, without reference to the underlying symmetry group.

The components of total angular momentum and total spin,
$L_\alpha=\sum_{j=1}^NL_\alpha(j)$ and $S_\alpha=\sum_{j=1}^NS_\alpha(j)$,
obey the usual commutator relations
$$
   [L_\alpha,L_\beta]=iL_\gamma, \;\;\; [S_\alpha,S_\beta]=iS_\gamma \;\;\;\; (\alpha,\beta,\gamma \mbox{ cyclic}).
$$
Angular momentum representation theory, together with
simple considerations concerning the above specific action of the operators on $N$-electron states,
yields the following well known facts (see e.g. \cite{Friesecke}).
\begin{lemma} \label{symlem} (a) For arbitrary $N$ and $Z$, a set of operators which commutes with the Hamiltonian $H$
and with each other is given by
\be \label{ops}
    \LL^2, \, L_3, \, \SSS^2, \, S_3, \, \Rhat.
\ee
(b) The eigenvalues of $\LL^2$, $\SSS^2$, and $\Rhat$ (acting on $L^2_a((\R^{3}\times\Z_2)^N)$) are, respectively,
\begin{align}
   & L(L+1), \;\; L=0,1,2,..., \\
   & S(S+1), \;\; S=\left\{\begin{array}{ll} \mbox{$\frac12$},\,\mbox{$\frac32$},\,\mbox{$\frac52$},...,\mbox{$\frac{N}{2}$}, & N \mbox{ odd}, \\
                                             0, \, 1, \, 2,...,\mbox{$\frac{N}{2}$}, & N \mbox{ even} \end{array}\right. \\
   & p=\pm 1.
\end{align}
(c) For fixed $L$, $S$ and $p$, on any joint eigenspace of $H$, $\LL^2$, $\SSS^2$ and $\Rhat$, $L_3$ has eigenvalues
$
               M=-L,-L+1,...,L,
$
and $S_3$ has eigenvalues
$
               M_S = -S, -S+1,...,S.
$
In particular, the eigenspace has dimension greater
or equal to $(2L+1)\cdot (2S+1)$, with equality in the case when the joint eigenspaces of $H$ and the operators (\ref{ops})
are non-degenerate (i.e., one-dimensional).
\end{lemma}
\noindent
Here we have employed the usual notation $\LL^2=L_1^2+L_2^2+L_3^2$ (and analogously for $\SSS^2$).

From the above we see that the main differences between the symmetries of many-electron atoms and those of hydrogen are the absence of an analogon of the
quantized Runge-Lenz vector and the nontrivial action of the spin operator
$\SSS^2$ (for $N=1$ it is equal to the trivial operator $\frac34 I$).
\\[2mm]
{\bf Definitions, 2.} The values of $L$, $S$ and $p$ for eigenstates of $\LL^2$, $\SSS^2$, $\Rhat$ are called the {\it total angular momentum quantum
number}, the {\it total spin quantum number}, and the {\it parity} of the state.
\\[2mm]
From the above lemma we see that for each energy level of $H$ there
exist unique quantum numbers $L$, $S$, $p$ characterizing the
symmetry of the eigenspace (except in ``non-generic'' cases where
the joint eigenspaces of $H$ and the operators (\ref{ops}) are
degenerate, in which case there exists a unique finite set of such
quantum numbers). In the chemistry literature these numbers for an
energy level are usually given in the form $^{2S+1}X^{\nu}$, where
$L$ corresponds to $X$ via $0\to S$, $1\to P$, $2\to D$, $3\to F$,
and where no superscript $\nu$ means $p=1$, and $\nu=o$ (for odd)
stands for $p=-1$. For example, the Carbon values $L=1$, $S=1$,
$p=1$ from Table \ref{Tab:GSqnumbers} would be denoted $^3P$, and
the Nitrogen values $L=0$, $S=3/2$, $p=-1$ by $^4S^o$.

Of particular physical and chemical interest is the energetic ordering in which different combinations of $L$ and $S$ appear in the spectrum of
the Hamiltonian (\ref{ham}); see Section \ref{Sec:AtSpec}.
\section{Ground states of the first ten atoms in the large $Z$ limit}
We are now in a position to state
a principal result of this paper.
\\[2mm]
{\bf Notation}
$|\eta_1\dots\eta_N\rangle$ denotes the Slater determinant (or antisymmetrized tensor product) of the orbitals
$\eta_j\!\in\! L^2(\R^3\times\Z_2)$,
$|\eta_1...\eta_N\rangle(x_1,s_1,..,x_N,s_N)={(N!)^{-1/2}}\det(\eta_i(x_j,s_j)_{i,j=1}^N)$.
$\phi\!\uparrow$, $\phi\!\downarrow$ stands for the spin-up and spin-down orbitals
$\phi(x)\delta_{\pm 1/2}(s)\in L^2(\R^3\times\Z_2)$. For a linear operator on the $N$-electron Hilbert space
(\ref{space}), $|||A|||$ denotes the usual operator norm $\sup\{||A\Psi|| \, : \, \Psi\in
L^2_a((\R^3\times\Z_2)^N), \, ||\Psi||=1\}$.
\begin{theorem} \label{GStheorem} Consider the atomic Schr\"odinger equation (\ref{SE}),
(\ref{ham}), (\ref{space}) for the atom/ion with $N$=1 to 10 electrons and nuclear charge $Z$. \\[1mm]
(i) For sufficiently large $Z$, the ground state has the spin and angular momentum quantum number $S$, $L$
and the dimension given in Table \ref{Tab:GSqnumbers}. \\[1mm]
(ii) In the limit $Z\to\infty$, the ground state is asymptotic to the explicit vector space given in Table
\ref{Tab:PTGS}, in the sense that the projection operators $P_0$, $\tilde{P}_0$ onto these spaces satisfy
$\lim_{Z\to\infty}|||P_0-\tilde{P}_0|||=0$.
Here $1$, $2$, $3$, $4$, $5$ stand for the scaled hydrogen orbitals (mathematically: hypergeometric functions)
$
    \phi_{1s}\!\!\uparrow, \;\phi_{2s}\!\!\uparrow, \;
    \phi_{2p_3}\!\!\uparrow, \; \phi_{2p_1}\!\!\uparrow, \; \phi_{2p_2}\!\!\uparrow
$
from (\ref{PTorbitals}), (\ref{porbitals}), and $\overline{1}$, $\overline{2}$, $\overline{3}$, $\overline{4}$, $\overline{5}$ denote the
corresponding spin-down orbitals.
\end{theorem}

\begin{table}[h!]
\resizebox{\textwidth}{!}{
\begin{tabular}{|c|c|c|c|}
\hline
$\mbox{Iso-electronic}\atop{\mbox{sequence}}$ & Symmetry & Ground state & Dimension \\
\hline
H  & $^2S$ & $|1\rangle$, $|\overline{1}\rangle$  & 2 \\

\hline He  & $^1S$ & $|1\overline{1}\rangle$ & 1 \\

\hline Li  & $^2S$ & $|1\overline{1}2\rangle$,
$|1\overline{1}\overline{2}\rangle$ & 2 \\

\hline Be  & $^1S$ & $\frac{1}{\sqrt{1+c^2}}\Big(
|1\overline{1}2\overline{2}\rangle + c \frac{1}{\sqrt{3}}\big(
|1\overline{1}3\overline{3}\rangle +
|1\overline{1}4\overline{4}\rangle +
|1\overline{1}5\overline{5}\rangle\big) \Big)$ & 1 \\
& & $c = - \frac{\sqrt{3}}{59049}(2\sqrt{1509308377} - 69821)
= -0.2310995\dots$ & \\

\hline B  & $^2P^o$ & $\frac{1}{\sqrt{1+c^2}}\Big(
|1\overline{1}2\overline{2}i\rangle + c \frac{1}{\sqrt{2}} \big(
|1\overline{1}ij\overline{j}\rangle +
|1\overline{1}ik\overline{k}\rangle \big) \Big)$ & 6 \\
& & $\frac{1}{\sqrt{1+c^2}}\Big(
|1\overline{1}2\overline{2}\overline{i}\rangle + c
\frac{1}{\sqrt{2}} \big(
|1\overline{1}\overline{i}j\overline{j}\rangle +
|1\overline{1}\overline{i}k\overline{k}\rangle \big) \Big)$ & \\
& & $(i,j,k)=(3,4,5),(4,5,3),(5,3,4)$ & \\
& & $c = -\frac{\sqrt{2}}{393660}(\sqrt{733174301809}-809747)
=-0.1670823\dots$ & \\

\hline C  & $^3P$ & $\frac{1}{\sqrt{1+c^2}}\big(
|1\overline{1}2\overline{2}ij\rangle + c
|1\overline{1}k\overline{k}ij\rangle \big)$ & 9 \\
& & $\frac{1}{\sqrt{1+c^2}}\Big( \frac{1}{\sqrt{2}}\big(
|1\overline{1}2\overline{2}i\overline{j}\rangle +
|1\overline{1}2\overline{2}\overline{i}j\rangle\big) +
c\frac{1}{\sqrt{2}}\big(
|1\overline{1}k\overline{k}i\overline{j}\rangle +
|1\overline{1}k\overline{k}\overline{i}j\rangle\big) \Big)$ & \\
& & $\frac{1}{\sqrt{1+c^2}}\big(
|1\overline{1}2\overline{2}\overline{i}\overline{j}\rangle + c
|1\overline{1}k\overline{k}\overline{i}\overline{j}\rangle \big)$ & \\
& & $(i,j,k)=(3,4,5),(4,5,3),(5,3,4)$ & \\
& & $c= - \frac{1}{98415}(\sqrt{221876564389}-460642)
=-0.1056317\dots$ & \\

\hline N  & $^4S^o$ & $|1\overline{1}2\overline{2}345\rangle$ & 4 \\
& &
$\frac{1}{\sqrt{3}}(|1\overline{1}2\overline{2}\overline{3}45\rangle
+ |1\overline{1}2\overline{2}3\overline{4}5\rangle +
|1\overline{1}2\overline{2}34\overline{5}\rangle)$ & \\
& &
$\frac{1}{\sqrt{3}}(|1\overline{1}2\overline{2}\overline{3}\overline{4}5\rangle
+ |1\overline{1}2\overline{2}\overline{3}4\overline{5}\rangle +
|1\overline{1}2\overline{2}3\overline{4}\overline{5}\rangle)$ & \\

& & $|1\overline{1}2\overline{2}\overline{3}\overline{4}\overline{5}\rangle$ & \\

\hline O  & $^3P$ &
$|1\overline{1}2\overline{2}i\overline{i}jk\rangle$ & 9 \\
& &
$\frac{1}{\sqrt{2}}(|1\overline{1}2\overline{2}i\overline{i}j\overline{k}\rangle
+ |1\overline{1}2\overline{2}i\overline{i}\overline{j}k\rangle)$ & \\
& & $|1\overline{1}2\overline{2}i\overline{i}\overline{j}\overline{k}\rangle$ & \\
& & $(i,j,k)=(3,4,5),(4,5,3),(5,3,4)$ & \\

\hline F  & $^2P^o$ &
$|1\overline{1}2\overline{2}i\overline{i}j\overline{j}k\rangle$ & 6 \\
& & $|1\overline{1}2\overline{2}i\overline{i}j\overline{j}\overline{k}\rangle$ & \\
& & $(i,j,k)=(3,4,5),(4,5,3),(5,3,4)$ & \\

\hline Ne  & $^1S$ &
$|1\overline{1}2\overline{2}3\overline{3}4\overline{4}5\overline{5}\rangle$ & 1 \\
\hline
\end{tabular}
} 
\caption{Ground states of the atomic Schr\"odinger equation in the limit $Z\to\infty$.
The indicated wave functions are an orthonormal basis of the ground
state. See Theorem \ref{GStheorem} for notation. The symmetry agrees with experiment
for each sequence and all $Z$.}\label{Tab:PTGS}
\end{table}

\noindent
In fact, all low-lying energy levels and eigenstates can be determined exactly in the above limit.
See Theorem \ref{mainresult}. Note also that, as we will see below,
the asymptotic ground states in the table are the exact ground states of the limit model (\ref{PT}),
(\ref{H0H'}), and become $Z$-independent after the re-scaling (\ref{scaling}).

The derivation of these results requires five steps: \\[1mm]
1. Reduction to the finite-dimensional problem $PHP\Psi=E\Psi$, $\Psi\in V_0(N)$ described in (\ref{PT}), (\ref{H0H'}). \\[1mm]
2. Explicit determination of $V_0(N)$. \\[1mm]
3. Choice of a suitable basis of $V_0(N)$ making $PHP$ particularly simple. \\[1mm]
4. Explicit determination of the $d\times d$ (i.e., in case of Carbon, $70\times 70$) matrix
representing the Hamiltonian $PHP$ from eq. (\ref{PT}) in this basis. \\[1mm]
5. Spectral analysis of this matrix.
\\[1mm]
These steps are carried out in Sections \ref{Sec:Pert}, \ref{Sec:GS}, \ref{S:Eigenspaces},
\ref{Sec:SymbolicVee}--\ref{Sec:ExplicitPHP}, and \ref{Sec:AtSpec}. Steps 1. and 2. follow from standard perturbation
theory respectively hydrogen atom theory.
3. is achieved by a basis adapted to the symmetries of $PHP$ (see Lemma \ref{symlem}) leading to block diagonal structure.
4. exploits, in addition, the fact that the Hamiltonian contains only one-body and two-body terms, allowing
to reduce evaluation of the required $N$-electron matrix elements $\langle\Psi|H|\tilde{\Psi}\rangle$, which are integrals
over $\R^{3N}$, to $1$- and $2$-electron matrix elements.

\subsection{Comparison with the semi-empirical Bohr-Hund-Slater picture of the periodic table}
\label{Sec:Aufbau} The result of Theorem \ref{GStheorem} provides a mathematical justification of the semi-empirical
``Aufbau principle'' (from the German word for building up)
developed notably by Bohr, Hund, and Slater to explain the periodic
table \cite{Bohr22, Hund25, LanLif, Schwabl01, Atkins01}. The aufbau principle is based on three semi-empirical postulates: \\[1mm]
(a) Each electron in an atom occupies a hydrogenic orbital.\footnote{In fact, in
Bohr's and Hund's original works \cite{Bohr22, Hund25}, which narrowly predate the Schr\"odinger equation,
the electrons were supposed to occupy hydrogenic Bohr orbits.}
\\[1mm]
(b) {\it Sub-shell ordering} The orbitals in each hydrogen energy level, or shell, form sub-shells which
are occupied in the order
$1s \;\;\; 2s \;\;\; 2p \;\;\; 3s \;\;\; 3p \;\;\; 4s \;\;\; 3d\;\;\; \dots$
\\[1mm]
(c) {\it Hund's rule} Within any partially filled sub-shell, the electrons adopt a configuration with the
greatest possible number of aligned spins.
\\[1mm]
Thus, for example, in Carbon the six electrons
would occupy the orbitals $1s\!\uparrow 1s\!\downarrow 2s\!\uparrow 2s\!\downarrow 2p_1\!\uparrow 2p_2\!\uparrow$ (note
that the alternative choices $2p_1\!\downarrow$ or $2p_2\!\downarrow$ for the last orbital would be consistent with (b)
but not (c)).

This beautiful heuristic picture is seen to emerge in Theorem \ref{GStheorem} in a natural way,
without reliance on the above nontrivial postulates or numerical simulations (up to small but interesting corrections).

(i) For seven out of ten elements (H, He, Li, N, O, F, Ne), the Aufbau principle configuration (when interpreted not as a collection of
individual electronic states, but as a Slater-determinantal many-electron wavefunction) is an element of the asymptotic
Schr\"odinger ground state.

(ii) For the remaining three elements (Be,
B, C), the Aufbau principle configuration is the dominant part of an element of the asymptotic Schr\"odinger ground state.

The corrections to the semi-empirical rules emerging in the large $Z$ limit of quantum mechanics are:
\\[1mm]
(1)
Alongside any Slater determinant, the asymptotic ground state contains
its orbit under the symmetry group of the many-electron Schr\"odin\-ger equation (see Lemma \ref{symlem}).
\\[1mm]
(2)
The corrections to the Aufbau principle configurations in (ii) come
from different sub-shells, indicating that rule (b)
is not strictly obeyed. We term this effect, which does not seem to have received
mathematical attention, {\bf 2s$^2$--2p$^2$ resonances}: besides the aufbau principle configurations
$1s^22s^2$, $1s^22s^22p$, and $1s^22s^22p^2$, a significant percentage is also present of,
respectively, $1s^22p^2$, $1s^22p^3$, and $1s^22p^4$. This could be
described as a resonance of the standard configuration with a
configuration in which the two 2s orbitals have been substituted by
two 2p orbitals.  Why this phenonenon occurs only for Be, B, C has a simple group-theoretic reason:
Tables \ref{Tab:VeeMatrix1}, \ref{Tab:VeeMatrix2} show that such a substitution
which preserves the total quantum numbers $L$ and $S$ is only possible in these three cases.
\\[1mm]
(3)
For excited states, Hund's rules are in rare cases found to disagree with the
experimental and mathematical results; see Section \ref{Sec:Comparison}.
\subsection{Comparison with other approaches} First we comment on the asymptotic regime in which the above picture emerges,
namely $N$ fixed, $Z\to\infty$. These parameters are the only ones contained in
the electronic Schr\"odinger equation that can be varied in ground states of a physical system. To capture chemical
specificity, $N$ must be kept fixed.
A priori, $Z$ could be made either small or large; but making it small leads to non-existence of bound electrons \cite{Lieb84}.
This leaves
$Z\to\infty$, i.e. the limit studied above, as the only option.

This limit has attracted considerable previous attention in the physics and chemistry literature in connection with
asymptotic expansions of energies and numerical evaluation of the coefficients. See Section 4 and the references given
there. But we do not know of any effort to derive a numerical analogue of Table \ref{Tab:PTGS} in this way (which
would correspond to truncating the expansion at first order in $1/Z$, see Theorem \ref{T:isolimit}), even though, in
principle, the tools to do so were available. In fact, in the only case we are aware of where first order wavefunctions
are given \cite[Table 1, p. 288]{Layzer59}, those for Be, B, C are incorrect
(the standard aufbau principle configurations are given, instead of the correlated states in Table \ref{Tab:PTGS}).

Another interesting strategy would be to vary mathematical parameters in the Schr\"odinger equation which cannot be varied physically, such as
$\hbar\to 0$ (semiclassical limit), or $D\to\infty$ where $D$ is the dimensionality of single-particle state space.
Despite interesting results (see e.g. the semiclassical picture of highly excited states of Helium in \cite{Tanner00}, and
the total atomic energies in \cite{Loeser87} via large-D asymptotics \cite{Witten80, Herschbach86}), these
ideas have so far not been aimed at, or led to, explanations of the main features of the periodic table.

The large majority of the literature on atomic systems is
computational, and does not take the Schr\"odinger equation, but
various simplified models as a starting point. Explanations of
the filling order (postulate (b) above) in terms of numerical solutions of the Hartree and Hartree-Fock equations began with the
pionieering work of Hartree on Rubidium \cite{Hartree28}; for treatments of large classes of atoms see e.g. \cite{Hartree57,
FroeseFischer77, Tatewaki94}. Note that these models assume a significant part of postulate (a) from the outset,
namely that electrons occupy individual orbitals and that these have $1s$, $2s$, $2p$, ... symmetry. The asymptotic Schr\"odinger
ground states which we determined above beautifully illustrate both the power  of the Slater determinant ansatz in
Hartree-Fock theory and its limitations:
in several cases other than noble gases the ground state contains a determinantal state, but for some of the atoms it
does not.

Finally we remark that the refined, multi-configurational methods developed in quantum chemistry (see e.g. \cite{FroeseFischer77,
Tayloretal1, SzaboOstlund96}) assume, on a somewhat ad hoc basis, exactly the structure of the wavefunctions
which emerges naturally in Table \ref{Tab:PTGS}, namely finite linear combinations of Slater determinants.
The only reason preventing these methods from being asymptotically exact within numerical error as $Z\to\infty$
is the use of Gaussians to represent the orbitals (see our companion paper \cite{FG09}).

\section{Reduction to Perturbation Theory Model}\label{Sec:Pert}

The first step in establishing the above result is to show that (\ref{SE}) simplifies
to a finite-dimensional model in the limit of fixed electron number $N$ and large nuclear charge $Z$.

If $\Psi$ solves the original Schr\"odinger equation (\ref{SE}),
then an elementary calculation shows that its rescaling
\be \label{scaling}
   \tilde{\Psi}(x_1,s_1,\dots,x_N,s_N) = Z^{-3N/2} \Psi(Z^{-1}x_1,s_1,\dots,Z^{-1}x_N,s_N)
\ee
solves the equation
\be \label{scaledSE}
  \Bigl(\tilde{H}_0 + \mbox{$\frac{1}{Z}$}V_{ee}\Bigr) \tilde{\Psi} = \Etilde \tilde{\Psi},
\ee
where $\tilde{H}_0$ is the $Z$-independent Hamiltonian
\be \label{ham0}
   \tilde{H}_0 = \sum_{i=1}^N\Bigl(-\frac12\Delta_{x_i}-\frac{1}{|x_i|}\Bigr)
\ee
(acting on the $N$-electron Hilbert space (\ref{space}))
and $\Etilde=\frac{1}{Z^2}E$. The elementary but important observation now is that the interaction term
$\frac{1}{Z}V_{ee}$ in (\ref{scaledSE}) becomes small if $Z$ is large, allowiong to treat the
interaction by perturbation theory.

Let us first derive the ensuing perturbation-theoretic model informally, then formulate a theorem.
By first order Rayleigh-Schr\"odinger perturbation theory, (\ref{scaledSE}) is expected to
be well approximated by
\begin{eqnarray} \label{scaledPTone}
   & & \tilde{P}(\tilde{H}_0 + \mbox{$\frac{1}{Z}$}V_{ee})\tilde{P}\tilde{\Psi} = \tilde{E}\tilde{\Psi}, \;\;
   \tilde{\Psi}\in \tilde{V}_0, \;\;\tilde{P}=\mbox{orth.projector onto }\tilde{V}_0 \\
   & & \tilde{V}_0 = \mbox{ground state eigenspace of }\tilde{H}_0, \label{scaledPTtwo}
\end{eqnarray}
with $\tilde{H}_0$ as in (\ref{ham0}). Now we undo the rescaling (\ref{scaling}). This yields the model
\begin{eqnarray} \label{PT}
   & & PHP\Psi = E\Psi, \;\;\;\Psi\in V_0, \;\;\; P=\mbox{orth.$\,$projector onto }V_0, \\
   & &
    V_0 = \mbox{ground state eigenspace of }H_0, \;\; H_0=\sum_{i=1}^N \Big( -\frac{1}{2} \Delta_{x_i} - \frac{Z}{|x_i|} \Big),
    \label{H0H'}
\end{eqnarray}
where $H$ is the original Hamiltonian (\ref{ham}).

We call eqns. (\ref{PT}), (\ref{H0H'}) the {\it PT model}. While it is still a
fully interacting quantum many-body model,
the key simplification is that the space $V_0$ is finite-dimensional. Its dimension for different atoms is easily read off
from Lemma \ref{L:meEfns} below:
\begin{table}[h!]
 \begin{center}
  \begin{tabular}{|r|c|c|c|c|c|c|c|c|c|}
   \hline
   Atom   & He& Li&Be  & B  & C  & N  & O  & F & Ne \\
   \hline
   $N$ & 2 & 3 & 4  & 5  & 6  & 7  & 8  & 9 & 10 \\
   dim $V_0$ & 1 & 8 & 28 & 56 & 70 & 56 & 28 & 8 & 1 \\
   \hline
  \end{tabular}
 \end{center}
 \caption
 {Dimensions of degenerate $H_0$ ground
 states, as given by Lemma \ref{L:meEfns}.}
 \label{Tab:H0Dim}
\end{table}

\noindent
An important feature of the PT model is that it
retains the full symmetries of the atomic Schr\"odinger equation.
\begin{lemma} \label{PTsymm} For arbitrary $N$ and $Z$, with $P$ as defined above and with $H$ denoting the
Hamiltonian (\ref{ham}), the operators (\ref{ops}) \\
(i) leave the ground state $V_0$ of $H_0$ invariant \\
(ii) commute with the PT Hamiltonian $PHP \, : \, V_0 \to V_0$.
\end{lemma}
\noindent
{\bf Proof} By direct inspection the operators (\ref{ops}) commute with $H_0$. Since $V_0$ is an eigenspace of
$H_0$, they must therefore leave $V_0$ invariant, and commute with the projector $P$ onto $V_0$.
As already shown (see Lemma \ref{symlem}), the operators (\ref{ops}) also commute with $H$,
and hence with the composition $PHP$.
\\[2mm]
We now come to the rigorous justification of the PT model (\ref{PT}), (\ref{H0H'}).
\begin{theorem} \label{T:isolimit} Let $N=1,\dots,10$, $Z>0$, and let $n(N)$ be the
number of energy levels of the PT model (\ref{PT}), (\ref{H0H'}). Then: \\[1mm]
(a) For all sufficiently large $Z$, the lowest $n(N)$ energy levels $E_1(N,Z)<\dots<E_{n(N)}(N,Z)$ of the
full Hamiltonian (\ref{ham}) have exactly the same dimension, total spin quantum number, total angular momentum
quantum number, and parity as the corresponding $PT$ energy levels $E_1^{PT}(N,Z)<\dots<E_{n(N)}^{PT}(N,Z)$. \\[1mm]
(b) The lowest $n(N)$ energy levels of the full Hamiltonian have the asymptotic expansion
$$
   \frac{E_j(N,Z)}{Z^2} = \frac{E_j^{PT}(N,Z)}{Z^2} + O(\mbox{$\frac{1}{Z^2}$}) =
   \Etilde^{(0)} + \frac{1}{Z}\Etilde^{(1)}_j + O(\mbox{$\frac{1}{Z^2}$}) \mbox{ as }Z\to\infty,
$$
where $\Etilde^{(0)}$ is the lowest eigenvalue of $\tilde{H}_0$
and the $\Etilde^{(1)}_j$ are the energy levels of $\tilde{P}V_{ee}\tilde{P}$ on $\tilde{V}_0$. \\[1mm]
(c) The projectors $P_1,\dots,P_{n(N)}$ onto the lowest $n(N)$ eigenspaces of the full Hamiltonian satisfy
$$
      |||P_j - P_j^{PT}|||\to 0 \mbox{ as }Z\to\infty,
$$
where the $P_j^{PT}$ are the corresponding projectors for the PT model.
\end{theorem}
\noindent
The idea that for large $Z$ the inter-electron term  $\Etilde^{(1)}_j$ provides the first order
correction to the non-interacting energy is well known in the physics literature  (see e.g.
\cite{Hylleraas30, BetheSalpeter57, SharmaCoulson62, SeungWilson67, RileyDalgarno71, Wilson84} who treat
non-degenerate eigenvalues, and see \cite{Layzer59} who gives an expansion similar to that in (b), not accounting
for multiplicities, and numerical tables of $\Etilde^{(1)}_j$ in the degenerate case). The main new insight here
is the absence of further splittings at higher orders of perturbation theory (see statement (a) in the theorem).
This is remarkable, considering that it fails in the simple $3\times 3$ matrix example
$$
   H(\epsilon) = \begin{pmatrix} 0 & 1 & 0 \\ 1 & 0 & 0 \\ 0 & 0 & 1\end{pmatrix}
          + \epsilon \begin{pmatrix} 1 & 0 & 0 \\ 0 & -1 & 0 \\ 0 & 0 & 0\end{pmatrix}.
$$
The eigenvalues are $\sqrt{1+\epsilon^2}$, $1$, $-\sqrt{1+\epsilon^2}$, and hence nondegenerate for $\epsilon\neq 0$,
but the leading eigenvalue is degenerate in first order perturbation theory.
\\[2mm]
{\bf Proof} Let $\tilde{E}_1^{PT}<...<\tilde{E}_{n(N)}^{PT}$ be the energy levels of the scaled model
(\ref{scaledPTone}), (\ref{scaledPTtwo}), and let $d_j$, $\tilde{P}_j^{PT}$ be the corresponding
eigenspace dimensions and eigenspace projectors. By perturbation theory for relatively bounded
perturbations of self-adjoint operators (see e.g. \cite{Kat95, Friesecke}), exactly $d_j$ eigenvalues of the scaled Schr\"odinger equation (\ref{scaledSE}) including multiplicity
are asymptotic to first order in $1/Z$ to the $j^{th}$ eigenvalue of (\ref{PT}), (\ref{H0H'}).
More precisely: The lowest $d_1+...+d_{n(N)}$ eigenvalues of
(\ref{scaledSE}) including multiplicity, labelled $\tilde{E}_{j,k}$, $j=1,..,n(N)$, $k=1,..,d_j$,
$\tilde{E}_{1,1}\le ... \le \tilde{E}_{1,d_1}\le \tilde{E}_{2,1}\le ... \le \tilde{E}_{2,d_2}\le ...$, satisfy
$$
   \tilde{E}_{j,k} = \tilde{E}^{(0)} + \frac{1}{Z}\tilde{E}_j^{(1)} + O\big(\frac{1}{Z^2}\big) \mbox{ as }Z\to\infty,
   \;\; k=1,..,d_j.
$$
Moreover the projector $\tilde{P}_j$ onto the span of these $d_j$ eigenstates satisfies
\be \label{threestar}
    ||| \tilde{P}_j - \tilde{P}_j^{PT} ||| \to 0 \mbox{ as }Z\to\infty.
\ee
Next, we investigate the Schr\"odinger eigenspace dimensions. By Lemma \ref{PTsymm}, each PT eigenspace possesses well defined spin, angular momentum and parity
quantum numbers $L$, $S$, and $p$, and by inspection of the explicit formulae in Theorem \ref{mainresult} below,
the space has minimal dimension subject to these numbers.
On the other hand, by (\ref{threestar}), for sufficiently large $Z$ these numbers must agree with those of the
eigenspaces of (\ref{scaledSE}); hence by Lemma \ref{symlem} (c), $\tilde{E}_{j,1}=...=\tilde{E}_{j,d_j}$.
Note that without the information on minimality of the PT dimensions, we would not be able to exclude
the possibility of further splittings of the Schr\"odinger eigenvalues beyond the PT splittings,
at higher orders of perturbation theory; this is the only reason why the restriction $N\le 10$ is needed.

The theorem now follows by applying the isometric scaling transformation (\ref{scaling}).

\section{State space of the PT model}\label{Sec:GS}
The important starting point for solving the PT model is the fact
that its state space, the GS of $H_0$, can be determined explicitly. This will follow
from the exact solubility of the Schr\"odinger equation of hydrogen and basic many-body arguments.
To explain these matters, we start from the hydrogen atom Hamiltonian
\be \label{hydrogen}
    h=-\tfrac{1}{2}\Delta - \frac{Z}{|x|},
\ee
$x\in\R^3$, acting on $L^2(\R^3\times\Z_2)$. For hydrogen, $Z=1$, but the parameter $Z>0$ will be useful later. Its
eigenvalues are given by (see e.g. \cite{Griffiths95})
\be \label{hydevals}
    e_n = - \frac{Z^2}{2n^2}, \;\;\; n\in\N,
\ee
and have corresponding $2n^2$-dimensional eigenspaces with orthonormal basis
\be \label{hydfctns}
    {\cal B}_n =\{ \phi_{n\ell m s}(x,s) \; | \; \ell=0, \dots , n-1, \; m=-\ell, -\ell+1, \dots,
    \ell, \; s=-\mbox{$\frac12$}, \mbox{$\frac12$} \},
\ee
where $\phi_{n \ell m s}\in L^2(\R^3\times\Z_2)$ is the, up to normalization, unique
eigenfunction of $h$, $\underline{L}^2$, $L_3$  and $S_3$ with
eigenvalues $-Z^2/(2n^2)$, $\ell(\ell+1)$, $m$ and $s$ respectively.

For later it will be useful to have an explicit form for these
so-called hydrogen orbitals, which in polar coordinates with spin coordinate $s\in\Z_2$ are
given by
\be
    \phi_{n,l,m,\sigma}(r,\theta,\phi,s) = \phi_{n\ell m}(r,\theta,\phi)\delta_\sigma(s) =
    Z^{3/2}R_{n,\ell}(Z r)Y_{\ell,m}(\theta,\phi)\delta_{\sigma}(s)
    \label{psinlm}
\ee
where
\be
    R_{n,\ell}(r):= \biggl( \Big(\frac{2}{n}\Big)^3
    \frac{(n-\ell-1)!}{2n[(n+\ell)!]} \biggr)^{1/2} e^{-r/n} \Big(
    \frac{2r}{n} \Big)^\ell
    L_{n-\ell-1}^{2\ell+1}\Big(\frac{2r}{n}\Big). \label{HRnl}
\ee
Here $L_n^k(x)$ is a generalized Laguerre polynomial and
$Y_{\ell,m}(\theta,\phi)$ is a spherical harmonic
\cite{AbramowitzStegun72}. In cartesian coordinates, the $n=1$ and $n=2$ orbitals are
\begin{align}
    \phi_{1,0,0}(x)&=\frac{Z^{3/2}}{\sqrt{\pi}} e^{-Z|x|} =:\phi_{1s}(x), \notag \\
    \phi_{2,0,0}(x)&=\frac{Z^{3/2}}{\sqrt{8 \pi}}
    \left(1-\frac{Z|x|}{2}\right)e^{-Z|x|/2} =:\phi_{2s}(x),  \notag \\
    \phi_{2,1,0}(x)&=\frac{Z^{5/2}}{\sqrt{32 \pi}} x_3
    e^{-Z\frac{|x|}{2}} =: \phi_{2p_3},
    \label{PTorbitals}\\
    \phi_{2,1,\pm 1}(x)&=\frac{Z^{5/2}}{\sqrt{32 \pi}} \frac{x_1\pm ix_2}{\sqrt{2}} e^{-Z|x|/2} =:\phi_{2p_\pm}. \notag
\end{align}
Often, it is convenient to work -- instead of the last two functions -- with their real linear
combinations
\be \label{porbitals}
    \frac{Z^{5/2}}{\sqrt{32 \pi}} x_j
    e^{-Z\frac{|x|}{2}} =: \phi_{2p_j}(x), \;\;\; j=1,2.
\ee

The following lemma describes how the eigenfunctions for the
non-interacting many-electron system are formed from these
one-electron eigenfunctions.
\begin{lemma} \label{L:meEfns} (Standard ``folklore'', see \cite{Friesecke} for a rigorous proof)
 \text{} (a)
      The lowest eigenvalue of the operator
      $$
                   H_0:=-\frac{1}{2}\sum_{i=1}^N \Delta_i - \sum_{i=1}^N\frac{Z}{|x_{i}|}
      $$
      on the space $L^2_a\big((\R^3\times\Z_2)^N\big)$ of square-integrable functions $\Psi : (\R^3 \times
      \Z_2)^N \rightarrow \C$ satisfying the antisymmetry condition (\ref{anti})
      is $E\!=\!\sum^N_{n=1}\tilde{e}_n$, where $\tilde{e}_1\!\le\!\tilde{e_2}\!\le\! ...$ is an ordered list including
      multiplicity of the hydrogen eigenvalues (\ref{hydevals}).
      \\[2mm]
      (b)
      The corresponding eigenspace is
      \begin{align}  V_0 = & \Span \Big\{|\chi_1\dots \chi_{d_*}\psi_{i_1} \dots
        \psi_{i_{N-d_*}} \rangle  \, \Big| \nonumber \\
        & \qquad\qquad 1\le i_1 < \dots <i_{N-d_*}\le 2(n_*+1)^2\}\Big\}, \label{manybodybasis}
      \end{align}
     where the functions $\chi_i$ and $\psi_i$ (``core orbitals'' and ``valence orbitals'') and the
     integers $d_*$ and $n_*$ (``number of core orbitals'' and ``number of closed shells'') are defined
     as follows:
     $d_*(N)$ is the largest number of form
     $\sum_{j=1}^n2j^2$ which is less or equal to $N$,
     $n_*(N)$ is the corresponding value of $n$,
     $$
         \{\chi_1,\dots,\chi_{d_*}\}={\cal B}_1\cup\cdots\cup{\cal B}_{n_*}
     $$
     (union of the ON bases (\ref{hydfctns}) of the first $n_*$ hydrogen eigenspaces), and
     $$
         \{\psi_1,\dots,\psi_{2(n_*+1)^2}\}={\cal B}_{n_*+1}
     $$
     (ON basis of the $(n_*+1)^{st}$ hydrogen eigenspace).
 \end{lemma}
Thus the ground state of the non-interacting
Hamiltonian is spanned by Slater determinants (alias antisymmetrized tensor products) formed
from scaled hydrogen orbitals, ``filled'' in order of increasing
one-electron energy.

Due to the freedom of choosing any $N-d_*$ eigenfunctions $\psi_i$ (``valence
orbitals'') from the basis of the highest relevant hydrogen eigenspace, whose dimension is $2(n_*+1)^2$, the noninteracting GS
has typically a large degeneracy:
\be
   d_0 := \mbox{dim GS of $H_0$} = \binom{2(n_*(N)+1)^2}{N-d_*(N)}. \label{dimGS}
\ee
{\bf Specialization to the second row atoms and their isoelectronic ions (N=3,...,10)} In this case, the number $d_*$ of
core orbitals equals $2$, the number $n_*$ of closed shells equals $1$,
and the dimension $2(n_*+1)^2$ of the hydrogen eigenspace from which the valence orbitals are selected
equals $8$. Thus by (\ref{dimGS}), the dimension of the ground state equals
$$
   d_0 = \binom{8}{N-2}.
$$
These numbers are given in Table \ref{Tab:H0Dim}. The set of core
respectively valence orbitals is (using the real orbitals
$\phi_{2p_1}$, $\phi_{2p_2}$ instead of $\phi_{2p_{\pm}}$)
\begin{align}
    & \{\chi_1,\chi_2\} = \{\phi_{1s}\!\uparrow , \, \phi_{1s}\!\downarrow \},  \label{Corbitals} \\
    & \{\psi_1,\dots,\psi_8\} = \{\phi_{2s}\!\uparrow, \, \phi_{2s}\!\downarrow, \,
                                      \phi_{2p_1}\!\uparrow, \, \phi_{2p_1} \!\downarrow, \,
                                      \phi_{2p_2}\!\uparrow, \, \phi_{2p_2}\!\downarrow, \,
                                      \phi_{2p_3}\!\uparrow, \, \phi_{2p_3} \!\downarrow \} \label{Vorbitals}.
\end{align}
Here we have employed the standard notation $\phi\!\uparrow$, $\phi\!\downarrow$ for the two spin orbitals
$\phi(x)\delta_{\pm 1/2}(\sigma)$.

Finally, the ground state of $H_0$ is
\be \label{GS2ndperiod}
  V_0(N) = \Span \Bigl\{ |\chi_1\chi_2\psi_{i_1}\dots\psi_{i_{N-2}}\rangle \, \Bigl|\Bigr. \, 1\le i_1<\dots<i_{N-2}\le 8\Bigr\}.
\ee

\section{Determining the matrix PHP} \label{Sec:Matrix}
In this section we determine explicitly the Hamiltonian matrices $PHP$, for all second period atoms.

Most of our arguments do not rely on the special radial form of the hydrogen orbitals
(\ref{Corbitals}), (\ref{Vorbitals}) appearing in the definition of the subspace $V_0(N)$. Hence in this section,
unless stated otherwise, $V_0(N)$ denotes the space
(\ref{GS2ndperiod}), (\ref{Corbitals}), (\ref{Vorbitals}) with the more general orbitals
\be
  \varphi_{1s}(x)=R_1(|x|), \;\; \varphi_{2s}(x)=R_2(|x|), \;\; \varphi_{2p_i}(x)=R_3(|x|)x_i \;(i=1,2,3), \label{GeneralOrbitals}
\ee
where the $\varphi$'s are in $L^2(\R^3)$ with norm one, $R_i: \R \to
\R$, and $\int_0^\infty R_1(r)R_2(r) r^2 dr=0$.

\subsection{Spin and angular momentum calculus on Slater determinants}\label{S:7.1}

The action of the spin and angular momentum operators on $V_0(N)$ can be calculated
from their action on the orbitals (\ref{Corbitals}) and (\ref{Vorbitals}) together
with the following simple identitites for the action of linear operators of form
$$
   B = \sum_{i=1}^N b(i), \;\;\; B^2 = \sum_{i,j=1}^N b(i) b(j)
$$
on Slater determinants, where $b$ is a linear operator on $L^2(\R^3\times\Z_2)$:
\begin{align}
   B \,  | \chi_1,\dots,\chi_N\rangle &= \sum_{i=1}^N |\chi_1, \dots, b\chi_i, \dots, \chi_N\rangle,
        \label{L:OneParticleOnSD} \\
   B^2 \,  | \chi_1,\dots,\chi_N\rangle &= \sum_{i=1}^N |\chi_1, \dots, b^2\chi_i, \dots, \chi_N\rangle \nonumber \\
         &\quad + 2 \sum_{1\le i<j\le N} |\chi_1,\dots, b\chi_i, \dots, b\chi_j, \dots, \chi_N\rangle.
        \label{L:TwoParticleOnSD}
\end{align}

Direct calculations show that, for any two spatial orbitals $\psi,
\phi \in L^2(\R^3)$, and orthogonal spin states $\alpha,
\beta:\Z_2\to\C$,
\begin{align*}
    \underline{S}\cdot \underline{S} \psi \alpha &= \tfrac{3}{4}
    \psi\alpha, \\
    (\underline{S}(1) \cdot \underline{S}(2)) \psi \alpha \otimes
    \phi \beta &= \tfrac{1}{2} \psi\beta \otimes \phi\alpha -
    \tfrac{1}{4} \psi\alpha \otimes \phi\beta, \\
    (\underline{S}(1) \cdot \underline{S}(2)) \psi\alpha \otimes \phi\alpha
     &= \tfrac{1}{4} \psi\alpha \otimes \phi\alpha.
\end{align*}

In particular, by (\ref{L:OneParticleOnSD}) and
(\ref{L:TwoParticleOnSD}), $S_3|\psi\alpha \psi\beta\rangle=0$ and
$\SSS^2|\psi\alpha \psi\beta\rangle = (
\frac{3}{4}+\frac{3}{4})|\psi\alpha \psi\beta\rangle - 2 \cdot
\frac{3}{4}|\psi\alpha \psi\beta\rangle=0$.

The angular momentum operators (\ref{Lj}) act on the orbitals (\ref{GeneralOrbitals}) as follows,
independently of the choice of spin $\alpha:\Z_2\to\C$:
\begin{align*}
   & L_j \varphi_{1s} \alpha = L_j \varphi_{2s} \alpha = L_j \varphi_{2p_j}\alpha =0, \\
     L_{j+1}\varphi_{2p_j}\alpha=&-i\varphi_{2p_{j-1}}\alpha,
    \quad L_{j-1}\varphi_{2p_j}\alpha=i\varphi_{2p_{j+1}}\alpha, \; j=1,2,3,
\end{align*}
where the indices are understood modulo three. Hence we need only
consider the action of $\LL^2$ on $\varphi_{2p_i}$, giving, for any
two spin states $\alpha$ and $\beta$, and $i \neq j$,
\begin{align*}
    \underline{L} \cdot \underline{L} \varphi_{2p_i} \alpha &= 2\varphi_{2p_i} \alpha \\
    (\underline{L}(1) \cdot \underline{L}(2)) \varphi_{2p_i}\alpha \otimes
    \varphi_{2p_i} \beta &= -(\varphi_{2p_{i-1}}\alpha \otimes \varphi_{2p_{i-1}}\beta + \varphi_{2p_{i+1}}\alpha
    \otimes \varphi_{2p_{i+1}} \beta) \\
    (\underline{L}(1) \cdot \underline{L}(2)) \varphi_{2p_i}\alpha \otimes
    \varphi_{2p_j} \beta &= \varphi_{2p_j}\alpha \otimes \varphi_{2p_i}\beta.
\end{align*}

Finally we see that, for any spin state $\alpha$,
$\hat{R}\varphi_{1s}\alpha=\varphi_{1s}\alpha$,
$\hat{R}\varphi_{2s}\alpha=\varphi_{2s}\alpha$ and
$\hat{R}\varphi_{2p_i}\alpha=-\varphi_{2p_i}\alpha$, $i=1,2,3$.

A useful and well known consequence of the above is that the pair of 1s orbitals makes no contribution
to spin, angular momentum, and parity on the space (\ref{GS2ndperiod}). More precisely:
\begin{lemma} \label{eliminate1s} (See [Fri0X])
The matrix of any of the operators (\ref{ops}) on $V_0(N)$ with respect
to the basis (\ref{GS2ndperiod}) is the same as that on the corresponding fewer-particle
space obtained by deleting the orbitals $\chi_1$, $\chi_2$,
with respect to the corresponding basis
$\{ | \psi_{i_1}\dots\psi_{i_{N-2}}\rangle \, | \, 1\le i_1<\dots<i_{N-2}\le
8\}$.
\end{lemma}

\subsection{Particle-hole duality}\label{S:Duality}

A further observation that simplifies the calculation of the eigenfunctions is a particle-hole duality result.
We introduce a
dual operator, in the spirit of the Hodge star operator from differential geometry (see e.g.
\cite{Jost02}), by
\\[2mm]
{\bf Definition} \label{D:star}
{\it The dual of $\alpha|\Psi\rangle$, with
$$
    |\Psi\rangle = |\varphi_{1s}\!\uparrow \varphi_{1s}\!\downarrow \psi_{i_1}
    \dots \psi_{i_{N-2}}\rangle,
$$
being any element of the basis (\ref{GS2ndperiod}) and
$\alpha \in \C$, denoted by $\ast(\alpha|\Psi\rangle)$, is given by
 \be
    \ast \big(\alpha|\Psi\rangle \big) := \alpha^* a(\psi_{i_{N-2}}) \dots
    a(\psi_{i_1})|\mathbbm{1}\rangle,
    \label{*2ndQ}
 \ee
where
$$
    |\mathbbm{1}\rangle:=|\varphi_{1s}\! \uparrow \varphi_{1s}\!\downarrow
    \varphi_{2s}\!\uparrow \varphi_{2s}\!\downarrow \varphi_{2p_1}\! \uparrow
    \varphi_{2p_1}\!\downarrow \varphi_{2p_2} \!\uparrow \varphi_{2p_2}\!\downarrow
    \varphi_{2p_3}\! \uparrow \varphi_{2p_3} \!\downarrow \rangle
$$
and $a(\psi)$ is the usual annihilation operator which maps $|\psi \, \psi_{i_1}\dots\psi_{i_k}\rangle$
to $|\psi_{i_1}\dots\psi_{i_k}\rangle$.
}
%
%
\\[2mm]
We extend $\ast$ linearly to real linear combinations of the
$\alpha|\Psi\rangle$, thereby obtaining an antilinear map from $V_0(N)$ to
$V_0(10-(N-2))$. We then have the following result:
\begin{lemma} \label{L:Duality}
Suppose $\Psi \in V_0(N)$ satisfies $\LL^2\Psi=\mathcal{L}\Psi$ and
$\SSS^2\Psi=\mathcal{S}\Psi$. Then
$\LL^2(\ast\Psi)=\mathcal{L}(\ast\Psi)$ and
$\SSS^2(\ast\Psi)=\mathcal{S}(\ast\Psi)$. Furthermore, if $L_3\Psi=M
\Psi$, $S_3 \Psi = M_s \Psi$ and $\hat{R}\Psi=p\Psi$, then
$L_3(\ast\Psi)=-M (\ast\Psi)$, $S_3 (\ast\Psi) = -M_s (\ast\Psi)$
and $\hat{R}(\ast\Psi)=p(\ast\Psi)$.
\end{lemma}
\noindent \textbf{Proof} Direct calculations using the second
quantized forms of $\LL$ and $\SSS$ show that both operators
anticommute with $\ast$ on $V_0(N)$. The results for angular momentum
and spin are then trivial.  The result for the inversion operator
follows from the fact that the parity of a wavefunction is
equivalent to the parity of the number of $p$-orbitals present in
each Slater determinant (since $\hat{R}\varphi_{ns} \alpha=
\varphi_{ns}\alpha$ for both $n=1$ and $2$, and
$\hat{R}\varphi_{2p_i}\alpha= -\varphi_{2p_i}\alpha$ for $i=1,2,3$)
and the number of $p$-orbitals in the dual of a Slater determinant
with $k$ $p$-orbitals is $6-k$, preserving the parity. \qed
\\[2mm]
The above result is a modest generalization of insights by spectroscopists
(who did not know they were speaking what mathematicians would call supersymmetry).
They introduced the dual of a configuration with respect to a single open shell
(termed ``conjugate configuration'' in \cite{Condon80}), and noticed that it gives rise to the same
set of $L_3$ and $S_3$ eigenvalues \cite{Condon80} and, what is more, the same
$\LL^2$, $\SSS^2$, and $\LL\cdot\SSS$ matrices \cite{CondonShortley35}.
\subsection{Simultaneous $\LL^2$-$\SSS^2$ Eigenspaces} \label{S:Eigenspaces}

We now form the joint angular momentum and spin eigenspaces within $V_0(N)$.
Lemma \ref{L:Duality} shows that we only need to do this for
Lithium-Carbon, the remaining cases follow by duality. Under the restriction of
a fixed number of $1s$, $2s$, $2p$ orbitals, the results were no doubt known to early
spectroscopists, who realized that the multiplet structure of observed spectra can only
be captured via superposition of aufbau principle Slater determinants into ``terms'' (in our
language, joint $\LL^2$--$\SSS^2$ eigenspaces); see e.g. \cite{CondonShortley35, Condon80}. We do not
however know of a complete tabulation.

\begin{theorem} For the Lithium-Neon sequences ($N=3, \dots, 10$, $Z>0$),
orthonormal basis for the $\LL^2$-$\SSS^2$ simultaneous eigenspaces
within $V_0(N)$ are as given in
Tables \ref{Tab:LiFullTable}-\ref{Tab:NeFullTable}. See below for
the notation used in the tables.
\end{theorem}
\noindent
\textbf{Proof} We only give the proof for the highest dimensional
case, Carbon, the other cases being analogous but easier.

By Lemma \ref{eliminate1s}, it suffices to find the joint $\LL^2$-$\SSS^2$-eigenstates
in the four-electron vector space spanned by
${\cal B}=\{|\psi_{i_1}\psi_{i_2}\psi_{i_3}\psi_{i_4}\rangle\, | \, 1\le i_1<i_2<i_3<i_4\le 8\}$,
with the $\psi_i$ as in (\ref{Vorbitals}).

We note first that each Slater determinant in the above basis
is already an eigenfunction of
$\SSS_3$, and that the space with $S_3$-eigenvalue $M$
is isomorphic (by flipping all spins) to that
with eigenvalue $-M$. Since both $\LL^2$ and $\SSS^2$ commute with $S_3$,
it suffices therefore to consider their action on the
eigenspaces of $S_3$ with eigenvalue $M\ge 0$.

Next we observe that within each such $S_3$-eigenspace, the span of
those Slater determinants which share the same number of different
spatial orbitals (4, 3, or 2) is also invariant under $\LL^2$ and
$\SSS^2$.

We now calculate the matrices of $\LL^2$ and $\SSS^2$ with respect to the
so-obtained subsets of the basis ${\cal B}$, using (\ref{L:OneParticleOnSD}), (\ref{L:TwoParticleOnSD}), and
the formulae from subsection \ref{S:7.1}.
To shorten the notation, we will write
$s\!\uparrow$, $s\!\downarrow$, $p_i\!\uparrow$, $p_i\!\downarrow$ instead of
$\varphi_{2s}\!\uparrow$, $\varphi_{2s}\!\downarrow$, $\varphi_{2p_i}\!\uparrow$, $\varphi_{2p_i}\!\downarrow$.
\\[1mm]
{\bf Four different spatial orbitals, M=2} On $|s\!\uparrow\, p_1\!\uparrow \, p_2\!\uparrow \,
p_3\!\uparrow\rangle$,
$$
  \LL^2 = 0, \;\;\; \SSS^2 = 6.
$$
{\bf Four different spatial orbitals, M=1} With respect to the basis
$\{|s\!\uparrow\, p_1\!\uparrow\, p_2\!\uparrow\, p_3\!\downarrow\rangle, \;
   |s\!\uparrow\, p_1\!\uparrow\, p_2\!\downarrow\, p_3\!\uparrow\rangle, \;
   |s\!\uparrow\, p_1\!\downarrow\, p_2\!\uparrow\, p_3\!\uparrow\rangle, \;
   |s\!\downarrow\, p_1\!\uparrow\, p_2\!\uparrow\, p_3\!\uparrow\rangle\}$,
$$
   \LL^2 = \left(\begin{array}{rrrr} 4 & -2 & -2 & 0 \\ -2 & 4 & -2 & 0 \\
                             -2 & -2 & 4 & 0 \\
                              0 &  0 & 0 & 0 \end{array}\right), \;\;\;
   \SSS^2 = \left(\begin{array}{rrrr} 3 & 1 & 1 & 1 \\ 1 & 3 & 1 & 1 \\
                              1 & 1 & 3 & 1 \\
                              1 & 1 & 1 & 3 \end{array}\right).
$$
{\bf Four different spatial orbitals, M=0} With respect to the basis
$\{|s\!\uparrow\, p_1\!\uparrow\, p_2\!\downarrow\, p_3\!\downarrow\rangle, \;
   |s\!\uparrow\, p_1\!\downarrow\, p_2\!\uparrow\, p_3\!\downarrow\rangle, \;
   |s\!\uparrow\, p_1\!\downarrow\, p_2\!\downarrow\, p_3\!\uparrow\rangle, \;
   |s\!\downarrow\, p_1\!\downarrow\, p_2\!\uparrow\, p_3\!\uparrow\rangle, \;
   |s\!\downarrow\, p_1\!\uparrow\, p_2\!\downarrow\, p_3\!\uparrow\rangle, \;
   |s\!\downarrow\, p_1\!\uparrow\, p_2\!\uparrow\, p_3\!\downarrow\rangle\}$,
$$
   \LL^2 = \left(\begin{array}{rrrrrr} 4 & -2 & -2 & 0 & 0 & 0 \\
                             -2 & 4 & -2 & 0 & 0 & 0 \\
                             -2 & -2 & 4 & 0 & 0 & 0 \\
                              0 &  0 & 0 & 4 & -2 & -2 \\
                              0 &  0 & 0 & -2 & 4 & -2 \\
                              0 &  0 & 0 & -2 & -2 & 4 \end{array}\right), \;\;\;
   \SSS^2 = \left(\begin{array}{rrrrrr} 2 & 1 & 1 & 0 & 1 & 1 \\
                              1 & 2 & 1 & 1 & 0 & 1 \\
                              1 & 1 & 2 & 1 & 1 & 0 \\
                              0 &  1 & 1 & 2 & 1 & 1 \\
                              1 &  0 & 1 & 1 & 2 & 1 \\
                              1 &  1 & 0 & 1 & 1 & 2 \end{array}\right). \;\;\;
$$
{\bf Three different spatial oritals, M=1} On each Slater determinant
$|s\!\uparrow\, s\!\downarrow\, p_i\!\uparrow\, p_j\!\uparrow\rangle$, and each Slater determinant
$|p_k\!\uparrow\, p_k\!\downarrow\, p_i\!\uparrow\, p_j\!\uparrow\rangle$,
$$
   \LL^2 = 2, \;\;\; \SSS^2=2.
$$
(In total, these span a 6-dimensional subspace.)
With respect to each basis
$\{|p_i\!\uparrow\, p_i\!\downarrow\, s\!\uparrow\, p_j\!\uparrow\rangle, \;
   |p_k\!\uparrow\, p_k\!\downarrow\, s\!\uparrow\, p_j\!\uparrow\rangle\}$,
$$
   \LL^2 = \left(\begin{array}{rr} 4 & -2 \\
                                  -2 &  4 \end{array}\right), \;\;\;
   \SSS^2 = 2.
$$
(In total, these span a 6-dimensional subspace.)
\\[1mm]
{\bf Three different spatial orbitals, M=0} With respect to each of the bases
$\{|s\!\uparrow\, s\!\downarrow\, p_i\!\uparrow\, p_j\!\downarrow\rangle, \;
   |s\!\uparrow\, s\!\downarrow\, p_i\!\downarrow\, p_j\!\uparrow\rangle\}$
and
$\{|p_k\!\uparrow\, p_k\!\downarrow\, p_i\!\uparrow\, p_j\!\downarrow\rangle, \;
   |p_k\!\uparrow\, p_k\!\downarrow\, p_i\!\downarrow\, p_j\!\uparrow\rangle\}$,
$$
   \LL^2 = \left(\begin{array}{rr} 4 & -2 \\
                                  -2 &  4 \end{array}\right), \;\;\;
   \SSS^2 = \left(\begin{array}{rr} 4 & -2 \\
                                  -2 &  4 \end{array}\right).
$$
(In total, these span a 12-dimensional subspace.)
With respect to each of the bases
$\{|p_i\!\uparrow\, p_i\!\downarrow\, s\!\uparrow\, p_j\!\downarrow\rangle, \;
   |p_k\!\uparrow\, p_k\!\downarrow\, s\!\uparrow\, p_j\!\downarrow\rangle, \;
   |p_i\!\uparrow\, p_i\!\downarrow\, s\!\downarrow\, p_j\!\uparrow\rangle, \;
   |p_k\!\uparrow\, p_k\!\downarrow\, s\!\downarrow\, p_j\!\uparrow\rangle$,
$$
   \LL^2 = \left(\begin{array}{rrrr} 4 & -2 & 0 & 0 \\
                                    -2 & 4 & 0 & 0 \\
                              0 & 0 & 4 & -2 \\
                              0 & 0 & -2 & 4 \end{array}\right), \;\;\;
   \SSS^2 = \left(\begin{array}{rrrr} 1 & 0 & 1 & 0 \\ 0 & 1 & 0 & 1 \\
                              1 & 0 & 1 & 0 \\
                              0 & 1 & 0 & 1 \end{array}\right).
$$
(In total, these span a 12-dimensional subspace.)
\\[1mm]
{\bf Two different spatial orbitals, M=0} With respect to the bases
$\{|s\!\uparrow\, s\!\downarrow\, p_1\!\uparrow\, p_1\!\downarrow\rangle, \;
   |s\!\uparrow\, s\!\downarrow\, p_2\!\uparrow\, p_2\!\downarrow\rangle, \;
   |s\!\uparrow\, s\!\downarrow\, p_3\!\uparrow\, p_3\!\downarrow\rangle\}$
and
$\{|p_1\!\uparrow\, p_1\!\downarrow\, p_2\!\uparrow\, p_2\!\downarrow\rangle, \;
   |p_2\!\uparrow\, p_2\!\downarrow\, p_3\!\uparrow\, p_3\!\downarrow\rangle, \;
   |p_3\!\uparrow\, p_3\!\downarrow\, p_1\!\uparrow\, p_1\!\downarrow\rangle\}$,
$$
   \LL^2 = \left(\begin{array}{rrr} 4 & -2 & -2  \\
                                   -2 & 4 & -2 \\
                                   -2 & -2 & 4 \end{array}\right), \;\;\;
   \SSS^2 = 0.
$$
This completes the explicit description of the action of $\LL^2$ and $\SSS^2$.

The eigenfunctions and eigenvalues are now found by explicit diagonalization of
the above matrices.                                            \qed\\

Tables \ref{Tab:LiFullTable}-\ref{Tab:NeFullTable} use the following
conventions: \\[1mm]
1) The two $1s$ orbitals present in every Slater
determinant are not shown. \\[1mm]
2) The eigenfunctions are not normalized. \\[1mm]
3) In all cases, it is assumed that $i = 1,2,3$,
$(i,j)=(1,2),(2,3),(3,1)$ and $(i,j,k)$ is any cyclic permutation of
$(1,2,3)$. In particular, any eigenfunction containing a variable corresponds to a 3D subspace. \\[1mm]
4) Eigenfunctions of the form $a\Psi_1 + b\Psi_2 +
c\Psi_3$ are such that $a+b+c=0$ and stand for two linearly
independent orthogonal choices of $(a,b,c)$, and hence correspond to a
two-dimensional subspace. \\[1mm]
5) Within each $\LL^2$-$\SSS^2$-eigenspace, the different
$S_3$ eigenspaces are separated by a line, in the order $M=S$, $-S$, $S-1$, $-(S-1)$, $\dots$, $0$. \\[1mm]
6) The spin orbitals
$\varphi_{2s}\!\uparrow$, $\varphi_{2s}\!\downarrow$, $\varphi_{2p_i}\!\uparrow$,
$\varphi_{2p_i}\!\downarrow$ are abbreviated $s$, $\overline{s}$, $p_i$,
$\overline{p_i}$. \\

Note that the parity of the eigenfunctions in the tables, although not shown explicitly,
can be read off by counting the number of $p$ orbitals (see the previous section).
%
%
%
%



\begin{table}[h!]
\begin{minipage}[h]{0.25\textwidth}

\begin{center}
\scalebox{0.75}{
  \begin{tabular}{|r|c|}
    \hline & $\SSS^2=\tfrac{3}{4}$ \\
    \hline

    $\LL^2=0$ &
    \begin{tabular}{c}
    $|s\rangle$ \\
    \hline
    $|\overline{s}\rangle$ \\
    \end{tabular} \\
    \hline

    $\LL^2=2$ &
    \begin{tabular}{c}
    $|p_i\rangle$ \\
    \hline
    $|\overline{p_i}\rangle$ \\
    \end{tabular} \\

    \hline
  \end{tabular}
} 
 \end{center}
 \caption{Lithium sequence $\protect\LL^2$-$\protect\SSS^2$
 eigenspaces.} \label{Tab:LiFullTable}
\end{minipage}
\begin{minipage}[h]{0.66\textwidth}

 \begin{center}

\scalebox{0.75}{
  \begin{tabular}{|r|c|c|}
    \hline & $\SSS^2=0$ & $\SSS^2=2$ \\

    \hline $\LL^2=0$ &

    \begin{tabular}{c}
    $|s\overline{s}\rangle$ \\
    $|p_1\overline{p_1}\rangle+|p_2\overline{p_2}\rangle+|p_3\overline{p_3}\rangle$\\
    \end{tabular}
    & \\

    \hline

    $\LL^2=2$

    &
    \begin{tabular}{c}
    $|s\overline{p_i}\rangle-|\overline{s}p_i\rangle$
    \end{tabular}

    &
    \begin{tabular}{cc}

    $|sp_i\rangle$

    & $|p_ip_j\rangle$\\
    \hline

    $|\overline{s}\overline{p_i}\rangle$

    & $|\overline{p_i}\overline{p_j}\rangle$ \\
    \hline

    $|s\overline{p_i}\rangle+|\overline{s}p_i\rangle$

    & $|p_i\overline{p_j}\rangle+|\overline{p_i}p_j\rangle$
    \vspace{1mm}
    \end{tabular}

    \\

    \hline

    $\LL^2=6$ &

    \begin{tabular}{c}
    $|p_i\overline{p_j}\rangle-|\overline{p_i}p_j\rangle$\\
    $a|p_1\overline{p_1}\rangle+b|p_2\overline{p_2}\rangle+c|p_3\overline{p_3}\rangle$
    \\
    \end{tabular}

    &\\

    \hline

  \end{tabular}

}

 \end{center}
 \caption{Beryllium sequence
 $\protect\LL^2$-$\protect\SSS^2$ eigenspaces.}
 \label{Tab:BeTableFull}
\end{minipage}

\end{table}

\begin{table}[h!tp]
 \begin{center}
\scalebox{0.75}{
  \begin{tabular}{|r|c|c|}
    \hline & $\SSS^2=\tfrac{3}{4}$
    & $\SSS^2=\tfrac{15}{4}$ \\

    \hline $\LL^2=0$ &

    \begin{tabular}{c}
    $|sp_1\overline{p_1}\rangle +|sp_2\overline{p_2}\rangle
    +|sp_3\overline{p_3}\rangle$ \\[3mm]
    \hline
    \rule{0pt}{6mm}

    $|\overline{s}p_1\overline{p_1}\rangle
    +|\overline{s}p_2\overline{p_2}\rangle
    + |\overline{s}p_3\overline{p_3}\rangle$\\
    \end{tabular}

    &

    \begin{tabular}{c}

    $|p_1p_2p_3\rangle$ \\
    \hline

    $|\overline{p_1}\overline{p_2}\overline{p_3}\rangle$ \\
    \hline

    $|p_1p_2\overline{p_3}\rangle+|p_1\overline{p_2}p_3\rangle+|\overline{p_1}p_2p_3\rangle$
    \\
    \hline

    $|\overline{p_1}\overline{p_2}p_3\rangle
    +|\overline{p_1}p_2\overline{p_3}\rangle
    +|p_1\overline{p_2}\overline{p_3}\rangle$
    \\

    \end{tabular}

    \\

    \hline $\LL^2=2$ &

    \begin{tabular}{c}

    $|s\overline{s}p_i\rangle$ \\

    $|p_ip_j\overline{p_j}\rangle
    +|p_ip_k\overline{p_k}\rangle$ \\

    $2|\overline{s}p_ip_j\rangle-|s\overline{p_i}p_j\rangle-|sp_i\overline{p_j}\rangle$\\
    \hline

    $|s\overline{s}\overline{p_i}\rangle$ \\

    $|\overline{p_i}p_j\overline{p_j}\rangle
    +|\overline{p_i}p_k\overline{p_k}\rangle$ \\

    $2|s\overline{p_i}\overline{p_j}\rangle-|\overline{s}p_i\overline{p_j}\rangle-|\overline{s}\overline{p_i}p_j\rangle$
    \\

    \end{tabular}

    &

    \begin{tabular}{c}

    $|sp_ip_j\rangle$ \\
    \hline

    $|\overline{s}\overline{p_i}\overline{p_j}\rangle$ \\
    \hline

    $|sp_i\overline{p_j}\rangle+|s\overline{p_i}p_j\rangle+|\overline{s}p_ip_j\rangle$\\
    \hline

    $|s\overline{p_i}\overline{p_j}\rangle+|\overline{s}p_i\overline{p_j}\rangle+|\overline{s}\overline{p_i}p_j\rangle$\\

    \end{tabular}

    \\

    \hline $\LL^2=6$ &

    \begin{tabular}{c}

    $|sp_i\overline{p_j}\rangle-|s\overline{p_i}p_j\rangle$ \\

    $a|sp_1\overline{p_1}\rangle
    +b|sp_2\overline{p_2}\rangle
    +c|sp_3\overline{p_3}\rangle$ \\

    $|p_ip_j\overline{p_j}\rangle-|p_ip_k\overline{p_k}\rangle$ \\

    $a|\overline{p_3}p_1p_2\rangle
    +b|p_3p_1\overline{p_2}\rangle
    +c|p_3\overline{p_1}p_2\rangle$ \\

    \hline

    $|\overline{s}p_i\overline{p_j}\rangle-|\overline{s}\overline{p_i}p_j\rangle$\\

    $a|\overline{s}p_1\overline{p_1}\rangle
    +b|\overline{s}p_2\overline{p_2}\rangle
    +c|\overline{s}p_3\overline{p_3}\rangle$
    \\

    $|\overline{p_i}p_j\overline{p_j}\rangle
    -|\overline{p_i}p_k\overline{p_k}\rangle$ \\

    $a|p_3\overline{p_1}\overline{p_2}\rangle
    +b|\overline{p_3}p_1\overline{p_2}\rangle
    +c|\overline{p_3}\overline{p_1}p_2\rangle$ \\

    \end{tabular}

    &
    \\
    \hline

  \end{tabular}
  } 
  \caption{Boron sequence $\protect\LL^2$-$\protect\SSS^2$
  eigenspaces.}
 \end{center}
\end{table}


\begin{table}[h!tp]
 \hspace{23mm}
 \rotatebox{90}{ %
 \resizebox{0.93\textheight}{!}{
  \begin{tabular}{|r|c|c|c|}
    \hline & $\SSS^2=0$ & $\SSS^2=2$ & $\SSS^2=6$ \\
    \hline
    $\LL^2=0$ &
    \begin{tabular}{c}
    $|s\overline{s}p_1\overline{p_1}\rangle
    +|s\overline{s}p_2\overline{p_2}\rangle
    +|s\overline{s}p_3\overline{p_3}\rangle$ \\
    $|p_1\overline{p_1}p_2\overline{p_2}\rangle
    +|p_1\overline{p_1}p_3\overline{p_3}\rangle
    +|p_2\overline{p_2}p_3\overline{p_3}\rangle$\\
    \end{tabular}
    &
    \begin{tabular}{c}
    $3|\overline{s}p_1p_2p_3\rangle -|sp_1p_2\overline{p_3}\rangle
    -|s\overline{p_1}p_2p_3\rangle -|sp_1\overline{p_2}p_3\rangle$\\[3mm]
    \hline
    $3|s\overline{p_1}\overline{p_2}\overline{p_3}\rangle
    -|\overline{s}\overline{p_1}\overline{p_2}p_3\rangle
    -|\overline{s}p_1\overline{p_2}\overline{p_3}\rangle
    -|\overline{s}\overline{p_1}p_2\overline{p_3}\rangle$
\rule{0pt}{6mm}\\[3mm]
    \hline
    $|sp_1\overline{p_2}\overline{p_3}\rangle +
    |s\overline{p_1}p_2\overline{p_3}\rangle +
    |s\overline{p_1}\overline{p_2}p_3\rangle$ \phantom{xxxxxx}\\
    \phantom{xxxxxx} $-|\overline{s}\overline{p_1}p_2p_3\rangle
    -|\overline{s}p_1\overline{p_2}p_3\rangle -
    |\overline{s}p_1p_2\overline{p_3}\rangle$ \rule{0pt}{6mm}\\
    \end{tabular}
    &
    \begin{tabular}{c}
    $|sp_1p_2p_3\rangle$\\
    \hline
    $|\overline{s}\overline{p_1}\overline{p_2}\overline{p_3}\rangle$\\
    \hline
    $|\overline{s}p_1p_2p_3\rangle +|sp_1p_2\overline{p_3}\rangle
    +|sp_1\overline{p_2}p_3\rangle+|s\overline{p_1}p_2p_3\rangle$\\
    \hline
    $|s\overline{p_1}\overline{p_2}\overline{p_3}\rangle +
    |\overline{s}\overline{p_1}\overline{p_2}p_3\rangle+
    |\overline{s}\overline{p_1}p_2\overline{p_3}\rangle+
    |\overline{s}p_1\overline{p_2}\overline{p_3}\rangle $\\
    \hline
    $|sp_1\overline{p_2}\overline{p_3}\rangle +
    |s\overline{p_1}p_2\overline{p_3}\rangle +
    |s\overline{p_1}\overline{p_2}p_3\rangle$ \phantom{xxxx}\\
    \phantom{xxxx}$+|\overline{s}\overline{p_1}p_2p_3\rangle
    +|\overline{s}p_1\overline{p_2}p_3\rangle +
    |\overline{s}p_1p_2\overline{p_3}\rangle$ \\
    \end{tabular}
    \\
    \hline
    $\LL^2=2$ &
    \begin{tabular}{c}
    $|s\overline{p_i}p_j\overline{p_j}\rangle -
    |\overline{s}p_ip_j\overline{p_j}\rangle +
    |s\overline{p_i}p_k\overline{p_k}\rangle -
    |\overline{s}p_ip_k\overline{p_k}\rangle$\\
    \end{tabular}
    &
    \begin{tabular}{c}
    $|s\overline{s}p_ip_j\rangle$\\
    $|p_1p_2p_3\overline{p_i}\rangle$ \\
    $|sp_ip_j\overline{p_j}\rangle + |sp_ip_k\overline{p_k}\rangle$
    \\
    \hline
    $|s\overline{s}\overline{p_i}\overline{p_j}\rangle$ \\
    $|\overline{p_1}\overline{p_2}\overline{p_3}p_i\rangle$\\
    $|\overline{s}\overline{p_i}p_j\overline{p_j}\rangle +
    |\overline{s}\overline{p_i}p_k\overline{p_k}\rangle$
    \\
    \hline
    $|s\overline{s}p_i\overline{p_j}\rangle +
    |s\overline{s}\overline{p_i}p_j\rangle$\\
    $|p_i\overline{p_i}p_j\overline{p_k}\rangle +
    |p_i\overline{p_i}\overline{p_j}p_k\rangle$ \\
    $|s\overline{p_i}p_j\overline{p_j}\rangle
    +|\overline{s}p_ip_j\overline{p_j}\rangle
    +|s\overline{p_i}p_k\overline{p_k}\rangle
    +|\overline{s}p_ip_k\overline{p_k}\rangle$\\
    \end{tabular}
    &
    \\\hline
    $\LL^2=6$ &
    \begin{tabular}{c}
    $|s\overline{s}p_i\overline{p_j}\rangle -
    |s\overline{s}\overline{p_i}p_j\rangle$\\
    $a|s\overline{s}p_1\overline{p_1}\rangle +
    b|s\overline{s}p_2\overline{p_2}\rangle +
    c|s\overline{s}p_3\overline{p_3}\rangle$\\
    $|p_i\overline{p_i}p_j\overline{p_k}\rangle -
    |p_i\overline{p_i}\overline{p_j}p_k\rangle$\\
    $a|p_1\overline{p_1}p_2\overline{p_2}\rangle +
    b|p_1\overline{p_1}p_3\overline{p_3}\rangle +
    c|p_2\overline{p_2}p_3\overline{p_3}\rangle$ \\
    $|s\overline{p_i}p_j\overline{p_j}\rangle-
    |\overline{s}p_ip_j\overline{p_j}\rangle
    -|s\overline{p_i}p_k\overline{p_k}\rangle +
    |\overline{s}p_ip_k\overline{p_k}\rangle$\\
    $a(|s\overline{p_1}\overline{p_2}p_3\rangle
    +|\overline{s}p_1p_2\overline{p_3}\rangle)
    +b(|sp_1\overline{p_2}\overline{p_3}\rangle +
    |\overline{s}\overline{p_1}p_2p_3\rangle)$\\
    $+c(|s\overline{p_1}p_2\overline{p_3}\rangle
    +|\overline{s}p_1\overline{p_2}p_3\rangle)$\\
    \end{tabular}
    &
    \begin{tabular}{c}
    $|sp_ip_j\overline{p_j}\rangle-|sp_ip_k\overline{p_k}\rangle$\\
    $a|sp_1p_2\overline{p_3}\rangle +b|sp_1\overline{p_2}p_3\rangle
    +c|s\overline{p_1}p_2p_3\rangle$\\
    \hline
    $|\overline{s}\overline{p_i}p_j\overline{p_j}\rangle
    -|\overline{s}\overline{p_i}p_k\overline{p_k}\rangle$\\
    $a|\overline{s}\overline{p_1}\overline{p_2}p_3\rangle
    +b|\overline{s}\overline{p_1}p_2\overline{p_3}\rangle
    +c|\overline{s}p_1\overline{p_2}\overline{p_3}\rangle$\\
    \hline
    $|s\overline{p_i}p_j\overline{p_j}\rangle+|\overline{s}p_ip_j\overline{p_j}\rangle
    -|s\overline{p_i}p_k\overline{p_k}\rangle-|\overline{s}p_ip_k\overline{p_k}\rangle$\\
    $a(|s\overline{p_1}\overline{p_2}p_3\rangle
    -|\overline{s}p_1p_2\overline{p_3}\rangle)
    +b(|sp_1\overline{p_2}\overline{p_3}\rangle -
    |\overline{s}\overline{p_1}p_2p_3\rangle)$\\
    $+c(|s\overline{p_1}p_2\overline{p_3}\rangle
    -|\overline{s}p_1\overline{p_2}p_3\rangle )$
    \end{tabular}
    & \\
    \hline
  \end{tabular}
 } 
 } 
 \caption{Carbon sequence $\protect\LL^2$-$\protect\SSS^2$
 eigenspaces.}
\end{table}

\begin{table}[h!tp]
 \begin{center}
  \resizebox{\textwidth}{!}{
   \begin{tabular}{|r|c|c|}
    \hline & $\SSS^2=\tfrac{3}{4}$ & $\SSS^2=\tfrac{15}{4}$ \\

    \hline $\LL^2=0$ &

    \begin{tabular}{c}
    $|sp_1\overline{p_1}p_2\overline{p_2}\rangle
    +|sp_1\overline{p_1}p_3\overline{p_3}\rangle
    +|sp_2\overline{p_2}p_3\overline{p_3}\rangle$\\[3mm]

    \hline \rule{0pt}{6mm}

    $|\overline{s}p_1\overline{p_1}p_2\overline{p_2}\rangle
    +|\overline{s}p_1\overline{p_1}p_3\overline{p_3}\rangle
    +|\overline{s}p_2\overline{p_2}p_3\overline{p_3}\rangle$\\
    \end{tabular}

    &

    \begin{tabular}{c}
    $|s\overline{s}p_1p_2p_3\rangle$\\
    \hline

    $|s\overline{s}\overline{p_1}\overline{p_2}\overline{p_3}\rangle$\\
    \hline

    $|s\overline{s}p_1\overline{p_2}p_3\rangle+|s\overline{s}\overline{p_1}p_2p_3\rangle+|s\overline{s}p_1p_2\overline{p_3}\rangle$
    \\ \hline

    $|s\overline{s}p_1\overline{p_2}\overline{p_3}\rangle+|s\overline{s}\overline{p_1}p_2\overline{p_3}\rangle+|s\overline{s}\overline{p_1}\overline{p_2}p_3\rangle$
    \\
    \end{tabular}

    \\

    \hline $\LL^2=2$ &

    \begin{tabular}{c}
    $|s\overline{s}p_ip_j\overline{p_j}\rangle
    +|s\overline{s}p_ip_k\overline{p_k}\rangle$\\

    $|p_1p_2p_3\overline{p_j}\overline{p_k}\rangle$ \\

    $2|\overline{s}p_i\overline{p_i}p_jp_k\rangle-|sp_i\overline{p_i}\overline{p_j}p_k\rangle
    -|sp_i\overline{p_i}p_j\overline{p_k}\rangle$\\

    \hline

    $|s\overline{s}\overline{p_i}p_j\overline{p_j}\rangle
    +|s\overline{s}\overline{p_i}p_k\overline{p_k}\rangle$\\

    $|\overline{p_1}\overline{p_2}\overline{p_3}p_jp_k\rangle$\\

    $2|sp_i\overline{p_i}\overline{p_j}\overline{p_k}\rangle
    -|\overline{s}p_i\overline{p_i}p_j\overline{p_k}\rangle
    -|\overline{s}p_i\overline{p_i}\overline{p_j}p_k\rangle$\\

    \end{tabular}

    &

    \begin{tabular}{c}
    $|sp_1p_2p_3\overline{p_i}\rangle$ \\ \hline

    $|\overline{s}\overline{p_1}\overline{p_2}\overline{p_3}p_i\rangle$
    \\ \hline

    $|sp_i\overline{p_i}p_j\overline{p_k}\rangle+
    |sp_i\overline{p_i}\overline{p_j}p_k\rangle
    +|\overline{s}p_i\overline{p_i}p_jp_k\rangle$ \\ \hline

    $|sp_i\overline{p_i}\overline{p_j}\overline{p_k}\rangle
    +|\overline{s}p_i\overline{p_i}p_j\overline{p_k}\rangle
    +|\overline{s}p_i\overline{p_i}\overline{p_j}p_k\rangle$\\
    \end{tabular}

    \\

    \hline $\LL^2=6$ &

    \begin{tabular}{c}

    $|s\overline{s}p_ip_j\overline{p_j}\rangle
    -|s\overline{s}p_ip_k\overline{p_k}\rangle$\\

    $a|s\overline{s}p_1p_2\overline{p_3}\rangle
    +b|s\overline{s}p_1\overline{p_2}p_3\rangle
    +c|s\overline{s}\overline{p_1}p_2p_3\rangle$\\

    $|sp_i\overline{p_i}p_j\overline{p_k}\rangle
    -|sp_i\overline{p_i}\overline{p_j}p_k\rangle$\\

    $a|sp_1\overline{p_1}p_2\overline{p_2}\rangle
    +b|sp_1\overline{p_1}p_3\overline{p_3}\rangle
    +c|sp_2\overline{p_2}p_3\overline{p_3}\rangle$\\

    \hline

    $|s\overline{s}\overline{p_i}p_j\overline{p_j}\rangle
    -|s\overline{s}\overline{p_i}p_k\overline{p_k}\rangle$\\

    $a|s\overline{s}\overline{p_1}\overline{p_2}p_3\rangle
    +b|s\overline{s}p_1\overline{p_2}\overline{p_3}\rangle
    +c|s\overline{s}\overline{p_1}p_2\overline{p_3}\rangle$\\

    $|\overline{s}p_i\overline{p_i}p_j\overline{p_k}\rangle
    -|\overline{s}p_i\overline{p_i}\overline{p_j}p_k\rangle$\\

    $a|\overline{s}p_1\overline{p_1}p_2\overline{p_2}\rangle
    +b|\overline{s}p_1\overline{p_1}p_3\overline{p_3}\rangle
    +c|\overline{s}p_2\overline{p_2}p_3\overline{p_3}\rangle$\\

    \end{tabular}

    &
    \\
    \hline

   \end{tabular}

  } 

 \end{center}
 \caption{Nitrogen sequence
 $\protect\LL^2$-$\protect\SSS^2$ eigenspaces.}
 \label{Tab:NFullTable}
\end{table}

\begin{table}[h!tp]
 \begin{center}
  \resizebox{\textwidth}{!}{
   \begin{tabular}{|r|c|c|}
    \hline & $\SSS^2=0$ & $\SSS^2=2$ \\

    \hline $\LL^2=0$ &

    \begin{tabular}{c}
    $|s\overline{s}p_1\overline{p_1}p_2\overline{p_2}\rangle
    +|s\overline{s}p_1\overline{p_1}p_3\overline{p_3}\rangle
    +|s\overline{s}p_2\overline{p_2}p_3\overline{p_3}\rangle$\\

    $|p_1\overline{p_1}p_2\overline{p_2}p_3\overline{p_3}\rangle$\\
    \end{tabular}
    & \\

    \hline

    $\LL^2=2$ &

    \begin{tabular}{c}
    $|s\overline{p_1}\overline{p_2}\overline{p_3}p_jp_k\rangle
    -|\overline{s}p_1p_2p_3\overline{p_j}\overline{p_k}\rangle$\\
    \end{tabular}

    &

    \begin{tabular}{c}
    $|s\overline{s}p_1p_2p_3\overline{p_i}\rangle$\\

    $|sp_1p_2p_3\overline{p_j}\overline{p_k}\rangle$\\

    \hline

    $|s\overline{s}\overline{p_1}\overline{p_2}\overline{p_3}p_i\rangle$\\

    $|\overline{s}\overline{p_1}\overline{p_2}\overline{p_3}p_jp_k\rangle$\\

    \hline

    $|s\overline{s}p_i\overline{p_i}p_j\overline{p_k}\rangle
    +|s\overline{s}p_i\overline{p_i}\overline{p_j}p_k\rangle$\\

    $|s\overline{p_1}\overline{p_2}\overline{p_3}p_jp_k\rangle
    +|\overline{s}p_1p_2p_3\overline{p_j}\overline{p_k}\rangle$\\

    \end{tabular}

    \\
    \hline

    $\LL^2=6$ &

    \begin{tabular}{c}
    $|s\overline{s}p_i\overline{p_i}p_j\overline{p_k}\rangle
    -|s\overline{s}p_i\overline{p_i}\overline{p_j}p_j\rangle$ \\

    $a|s\overline{s}p_1\overline{p_1}p_2\overline{p_2}\rangle
    +b|s\overline{s}p_1\overline{p_1}p_3\overline{p_3}\rangle
    +c|s\overline{s}p_2\overline{p_2}p_3\overline{p_3}\rangle$ \\
    \end{tabular}
     & \\

    \hline

   \end{tabular}
  } 
 \end{center}

 \caption{Oxygen sequence $\protect\LL^2$-$\protect\SSS^2$
 eigenspaces.}
\end{table}

\begin{table}[h!tp]
\begin{minipage}{0.5\textwidth}
 \begin{center}
  \begin{tabular}{|r|c|}
    \hline & $\SSS^2=\tfrac{3}{4}$ \\

    \hline

    $\LL^2=0$ &

    \begin{tabular}{c}
    $|sp_1\overline{p_1}p_2\overline{p_2}p_3\overline{p_3}\rangle$
    \\
    \hline

    $|\overline{s}p_1\overline{p_1}p_2\overline{p_2}p_3\overline{p_3}\rangle$\\
    \end{tabular}

    \\
    \hline

    $\LL^2=2$ &

    \begin{tabular}{c}
    $|s\overline{s}p_1p_2p_3\overline{p_i}\overline{p_j}\rangle$\\
    \hline

    $|s\overline{s}\overline{p_1}\overline{p_2}\overline{p_3}p_ip_j\rangle$\\
    \end{tabular}
    \\
    \hline

  \end{tabular}
 \end{center}
 \caption{Fluorine sequence
 $\protect\LL^2$-$\protect\SSS^2$ eigenspaces.}
 \label{Tab:FFullTable}
\end{minipage}
\begin{minipage}{0.5\textwidth}
 \begin{center}
  \begin{tabular}{|r|c|}
    \hline & $\SSS^2=0$ \\
    \hline

    $\LL^2=0$ & $|s\overline{s}p_1\overline{p_1}p_2\overline{p_2}p_3\overline{p_3}\rangle$ \\
    \hline

  \end{tabular}
 \end{center}
 \caption{Neon sequence $\protect\LL^2$-$\protect\SSS^2$
 eigenspaces.} \label{Tab:NeFullTable}
\end{minipage}
\end{table}

Inspecting these eigenspaces reveals a number of interesting properties.
\begin{corollary} For any $N=3,\dots,10$, the maximum dimension of any simultaneous eigenspace within $V_0(N)$ of
the operators (\ref{ops}) is two.
\end{corollary}
As regards diagonalization of the Hamiltonian,
this is clearly much more promising
than the 70-dimensional space of Carbon.
\begin{corollary}
\label{C:SDDifferByEven} For any $N=3,\dots,10$, and any simultaneous
$\LL^2$-$\SSS^2$-$L_3$-$S_3$-$\hat{R}$ eigenspace within $V_0(N)$ with $L_3$ eigenvalue equal to zero, all Slater
determinants occuring within the space differ by an even
number of orbitals.
\end{corollary}
This is remarkable, and will greatly simplify the structure of the
Hamiltonian matrix in the basis (\ref{GS2ndperiod}), due to
the simpler structure of Slater's rules (see below). Also, it
implies that even for the correlated eigenstates of the Hamiltonian,
the orbitals (\ref{Vorbitals}) are natural orbitals in the sense of
L\"owdin. A more abstract proof of
Corollary \ref{C:SDDifferByEven} will be given elsewhere.
\subsection{Symbolic interaction matrix}\label{Sec:SymbolicVee}
In order to calculate the Hamiltonian matrix $PHP$ on each $\LL^2$-$\SSS^2$-$\hat{R}$ eigenspace, note first
that $PHP$ commutes with (\ref{ops}) (see Lemma \ref{PTsymm}, which remains valied for the more general orbitals (\ref{GeneralOrbitals}), cf.
the calculus in Section \ref{S:7.1}). Hence it suffices to pick arbitrary components of $\LL$ and $\SSS$, say $L_3$ and $S_3$, and
calculate this matrix on the $\LL^2$-$L_3$-$\SSS^2$-$S_3$-$\hat{R}$ eigenspace with maximal $S_3$ and $L_3=0$.
These spaces are shown in Tables \ref{Tab:VeeMatrix1}--\ref{Tab:VeeMatrix3}.
Here, because of their importance for the interaction energy, the
$1s$ orbitals are shown and the eigenfunctions are normalized. We
find it convenient to abbreviate the spin orbitals
\be \label{not1}
   \varphi_{1s}\!\uparrow,\;\varphi_{1s}\!\downarrow,\;\varphi_{2s}\!\uparrow, \;
   \varphi_{2s}\!\downarrow,\;\varphi_{2p_3}\!\uparrow,\;\varphi_{2p_3}\!\downarrow, \;
   \varphi_{2p_1}\!\uparrow,\;\varphi_{2p_1}\!\downarrow, \;
   \varphi_{2p_2}\!\uparrow,\;\varphi_{2p_2}\!\downarrow
\ee
(even more drastically than in Tables
\ref{Tab:LiFullTable}--\ref{Tab:NeFullTable}) by
\be \label{not2}
   1,\;\overline{1},\;2,\;\overline{2},\;3,\;\overline{3},\;4,\;\overline{4},\;5,\;\overline{5}.
\ee
Thus, for example, the top Carbon state of Table \ref{Tab:VeeMatrix2},
$$
  \frac{1}{\sqrt{3}}( |1\overline{1}2\overline{2}3\overline{3}\rangle
  +|1\overline{1}2\overline{2}4\overline{4}\rangle +
  |1\overline{1}2\overline{2}5\overline{5}\rangle),
$$
stands for
\begin{align*}
 \frac{1}{\sqrt{3}}\Bigl( |\varphi_{1s}\!\!\uparrow \varphi_{1s}\!\!\downarrow
\varphi_{2s}\!\!\uparrow \varphi_{2s}\!\!\downarrow \varphi_{2p_3}\!\!\uparrow
\varphi_{2p_3}\!\!\downarrow \rangle  & + |\varphi_{1s}\!\!\uparrow
\varphi_{1s}\!\!\downarrow \varphi_{2s}\!\!\uparrow \varphi_{2s}\!\!\downarrow
\varphi_{2p_1}\!\!\uparrow \varphi_{2p_1}\!\!\downarrow \rangle \\
                                   & +
|\varphi_{1s}\!\!\uparrow \varphi_{1s}\!\!\downarrow \varphi_{2s}\!\!\uparrow
\varphi_{2s}\!\!\downarrow \varphi_{2p_2}\!\!\uparrow \varphi_{2p_2}\!\!\downarrow
\rangle\Bigr).
\end{align*}

We begin by analyzing the $V_{ee}$ matrix elements between
the eigenfunctions of Tables
\ref{Tab:VeeMatrix1}--\ref{Tab:VeeMatrix3}. Using Slater's rules
\cite[Section 2.3]{SzaboOstlund96}, these are straightforward to
express in terms of Coulomb and exchange integrals $(aa|bb)$ and
$(ab|ba)$ of the spatial orbitals (\ref{GeneralOrbitals}), where -- in common notation --
\be
    (ab | cd)
    = \int_{\R^6} dx_1 dx_2 a^*(x_1)b(x_1)
    \frac{1}{|x_1-x_2|} c^*(x_2)d(x_2).
    \label{ijklIntegral}
\ee

\begin{lemma} \label{L:7.3}
    Let $N\in\{3,\dots,10\}$.
    Orthonormal bases of the simultaneous $\LL^2$-$\SSS^2$-$L_3$-$S_3$-$\hat{R}$
    eigenspaces within $V_0(N)$ with $S_3$ maximal and $L_3=0$, and the corresponding
    $V_{ee}$ matrix elements $\langle\Psi|V_{ee}|\tilde{\Psi}\rangle$ in terms of
    Coulomb and exchange integrals of the one-electron
    orbitals (\ref{GeneralOrbitals}), are as given in Tables
    \ref{Tab:VeeMatrix1}-\ref{Tab:VeeMatrix3}. Here the orbitals are abbreviated as in (\ref{not1})--(\ref{not2}),
    and the off-diagonal matrix elements ($\Psi\neq\tilde{\Psi}$) in the
    two-dimensional eigenspaces are denoted by ``cross''.
\end{lemma}

The shortness of the expressions for the $\langle\Psi|V_{ee}|\tilde{\Psi}\rangle$, and the absence of
Coulomb and exchange integrals involving the last orbital, comes from
the absence of single excitations (Cor.~\ref{C:SDDifferByEven}) and the equivalence of the $p$ orbitals
in (\ref{GeneralOrbitals}) up to rotation. The latter would be destroyed by changing to a basis of $L_3$ eigenfunctions
(which is why we have not done so even though this would have been more convenient
for the diagonalization of $\LL^2$ in the previous subsection).

\begin{table}[htbp]
 \begin{center}

  \resizebox{\textwidth}{!}{
   \begin{tabular}{|c|c|c|c|c|l|}
    \hline & $\LL^2$&
    $\SSS^2$ &  $\hat{R}$& $\Psi$
    & \multicolumn{1}{c|}{$\langle V_{ee}\rangle$} \\
    \hline

    Li & $0$ & $\tfrac{3}{4}$ & $1$ & $|1\overline{1}2\rangle$ & $(11|11) + 2(11|22) - (12|21)$\\

    \cline{2-6} & $2$ & $\tfrac{3}{4}$ & $-1$  & $|1\overline{1}3\rangle$ &
    $(11|11) + 2(11|33)-(13|31)$
    \\
    \hline

    Be & $0$ & $0$ & $1$ & $|1\overline{1}2\overline{2}\rangle$ &$(11|11) +
    4(11|22) - 2(12|21) + (22|22)$\\

    \cline{5-6} & & & &
    $\tfrac{1}{\sqrt{3}}\left(|1\overline{1}3\overline{3}\rangle +
    |1\overline{1}4\overline{4}\rangle +
    |1\overline{1}5\overline{5}\rangle \right)$ & $(11|11) + 4(11|33)
    - 2(13|31) + (33|33) + 2(34|43)$ \\

    \cline{5-6} & & & & cross &
     $\sqrt{3}(23|32)$ \\

    \cline{2-6} & $2$ & $0$ & $-1$ &
    $\tfrac{1}{\sqrt{2}}\left(|1\overline{1}2\overline{3}\rangle
    -|1\overline{1}\overline{2}3\rangle\right)$ &  $(11|11) + 2(11|22)
    - (12|21) +2(11|33)- (13|31)$ \\
    & & & & & $ + (22|33) + (23|32)$ \\

    \cline{2-6} & $2$ & $2$ & $-1$ & $|1\overline{1}23\rangle$ & $(11|11) +
    2(11|22) - (12|21) + 2(11|33)- (13|31)$ \\
    & & & & & $ + (22|33) - (23|32)$\\

    \cline{4-6} & & & $1$ & $|1\overline{1}45\rangle$ &$(11|11) + 4(11|33)
    -2(13|31) + (33|44) - (34|43)$ \\

    \cline{2-6} & $6$ & $0$ & $1$ &
    $\tfrac{1}{\sqrt{6}}\left(2|1\overline{1}3\overline{3}\rangle-
    |1\overline{1}4\overline{4}\rangle
    -|1\overline{1}5\overline{5}\rangle \right)$
     &$(11|11) + 4(11|33)
    -2(13|31) + (33|33) -(34|43)$\\

    \hline B & $0$ & $\tfrac{3}{4}$ & $1$ &
    $\tfrac{1}{\sqrt{3}}\left(|1\overline{1}23\overline{3}\rangle +
    |1\overline{1}24\overline{4}\rangle +
    |1\overline{1}25\overline{5}\rangle\right)$&
     $(11|11) + 2(11|22) - (12|21) + 4(11|33) -2(13|31)$ \\
    & & & & &  $+ 2(22|33) - (23|32) + (33|33) + 2(34|43)$\\

    \cline{2-6} & $0$ & $\tfrac{15}{4}$ & $-1$ & $|1\overline{1}345\rangle$ &
    $(11|11) + 6(11|33) -3(13|31) + 3(33|44) -3(34|43)$ \\

    \cline{2-6} & $2$ & $\tfrac{3}{4}$ & $-1$ &
    $|1\overline{1}2\overline{2}{3}\rangle$ &
     $(11|11) + 4(11|22) -2(12|21) + 2(11|33) - (13|31) $\\
    & & & & & $ + (22|22) + 2(22|33) - (23|32)$ \\

    \cline{5-6} & & & &
    $\tfrac{1}{\sqrt{2}}\left(|1\overline{1}34\overline{4}\rangle +
    |1\overline{1}35\overline{5}\rangle \right)$ & $ (11|11) +
    6(11|33) -3(13|31) + (33|33) + 2(33|44)$ \\

    \cline{5-6} & & & & cross & $\sqrt{2}(23|32)$\\

    \cline{4-6} & & & $1$ &
    $\tfrac{1}{\sqrt{6}}\left(2|1\overline{1}\overline{2}45\rangle
    -|1\overline{1}2\overline{4}5\rangle
    -|1\overline{1}24\overline{5}\rangle \right)$
    & $(11|11) + 2(11|22) -(12|21) + 4(11|33) -2(13|31) $ \\
    & & & & & $+ 2(22|33) + (23|32) + (33|44) -(34|43) $ \\

    \cline{2-6} & $2$ & $\tfrac{15}{4}$ & $1$ & $|1\overline{1}245\rangle$ &
    $(11|11) + 2(11|22) - (12|21) + 4(11|33) -2(13|31)$ \\
    & & & & & $+ 2(22|33) -2(23|32) + (33|44) - (34|43)$\\

    \cline{2-6} & $6$ & $\tfrac{3}{4}$ & $1$ &
    $\tfrac{1}{\sqrt{6}}\left(2|1\overline{1}23\overline{3}\rangle
    -|1\overline{1}24\overline{4}\rangle
    -|1\overline{1}25\overline{5}\rangle \right)$
    & $(11|11) + 2(11|22) -(12|21) + 4(11|33) -2(13|31)$ \\
    & & & & & $ + 2(22|33) -(23|32) + (33|33) -(34|43)$ \\

    \cline{4-6} & & & $-1$ &
    $\tfrac{1}{\sqrt{6}}\left(2|1\overline{1}\overline{3}45\rangle
    -|1\overline{1}3\overline{4}5\rangle
    -|1\overline{1}34\overline{5}\rangle \right)$ & $(11|11) +
    6(11|33) -3(13|31) + 3(33|44)$ \\
    \hline

   \end{tabular}
  } 
 \end{center}
 \caption[$V_{ee}$ matrix element expressions for Li-B]
 {$V_{ee}$ matrix element expressions for the
 Li-B sequences, `cross' denotes the
 off-diagonal term in the $2\times2$ matrix. See Lemma \ref{L:7.3} for notation.}
 \label{Tab:VeeMatrix1}
\end{table}

\begin{table}[htbp]
 \begin{center}

  \resizebox{\textwidth}{!}{
   \begin{tabular}{|c|c|c|c|c|l|}
    \hline & $\LL^2$&
    $\SSS^2$ &  $\hat{R}$& $\Psi$
    & \multicolumn{1}{c|}{$\langle V_{ee} \rangle$} \\
    \hline

    C & $0$ & $0$ & $1$ &
    $\tfrac{1}{\sqrt{3}}\left(|1\overline{1}2\overline{2}3\overline{3}\rangle
    + |1\overline{1}2\overline{2}4\overline{4}\rangle +
    |1\overline{1}2\overline{2}5\overline{5}\rangle \right) $
    & $(11|11) + 4(11|22) -2(12|21) + 4(11|33) -2(13|31)$ \\
    & & & & & $+ (22|22) + 4(22|33) -2(23|32) + (33|33) + 2(34|43)$\\

    \cline{5-6} & & & &
    $\tfrac{1}{\sqrt{3}}\left(|1\overline{1}3\overline{3}4\overline{4}\rangle
    + |1\overline{1}3\overline{3}5\overline{5}\rangle +
    |1\overline{1}4\overline{4}5\overline{5}\rangle \right) $
    & $(11|11) + 8(11|33) -4(13|31) + 2(33|33) + 4(33|44)$ \\

    \cline{5-6} & & & &
    cross
    &$2(23|32)$ \\

    \cline{2-6} & $0$ & $2$ & $-1$ & $\hspace{-15mm} \tfrac{1}{\sqrt{12}}\big(
    3|1\overline{1}\overline{2}345\rangle -
    |1\overline{1}2\overline{3}45\rangle $
    & $(11|11) + 2(11|22) - (12|21) + 6(11|33) -3(13|31) $ \\
    & & & & $\hspace{15mm} -|1\overline{1}23\overline{4}5\rangle-
    |1\overline{1}234\overline{5}\rangle \big)$ &
    $+ 3(22|33) + (23|32) + 3(33|44) -3(34|43) $ \\

    \cline{2-6} & $0$ & $6$ & $-1$ & $|1\overline{1}2345\rangle$ & $(11|11) +
    2(11|22) - (12|21) + 6(11|33) -3(13|31)$\\
    & & & & &  $+ 3(22|33) -3(23|32) + 3(33|44) -3(34|43)$ \\

    \cline{2-6} & $2$ & $0$ & $-1$ &
    $\hspace{-15mm} \tfrac{1}{2}\big(|1\overline{1}2\overline{3}4\overline{4}\rangle -
    |1\overline{1}\overline{2}34\overline{4}\rangle$
    & $(11|11) + 2(11|22) - (12|21) + 6(11|33) -3(13|31)$ \\
    & & & & $\hspace{15mm} +|1\overline{1}2\overline{3}5\overline{5}\rangle
    - |1\overline{1}\overline{2}35\overline{5}\rangle \big)$
    & $+ 3(22|33) + (33|33) + 2(33|44)$ \\

    \cline{2-6} & $2$ & $2$ & $1$ & $|1\overline{1}2\overline{2}45\rangle$
    &  $(11|11) + 4(11|22) -2(12|21) + 4(11|33) -2(13|31)$  \\
    & & & & & $ + (22|22) + 4(22|33) -2(23|32) + (33|44) - (34|43)$ \\

    \cline{5-6} & & & & $|1\overline{1}3\overline{3}45\rangle$ &
    $(11|11) + 8(11|33) -4(13|31) + (33|33) + 5(33|44)$ \\
    & & & & & $-3(34|43)$\\

    \cline{5-6} & & & & cross & $(23|32)$ \\

    \cline{4-6} & & & $-1$ &
    $\tfrac{1}{\sqrt{2}}\left(|1\overline{1}234\overline{4}\rangle +
    |1\overline{1}235\overline{5}\rangle \right)$
    & $(11|11) + 2(11|22) -(12|21) + 6(11|33) -3(13|31)$ \\
    & & & & & $+ 3(22|33) -2(23|32) + (33|33) + 2(33|44)$ \\

    \cline{2-6} & $6$ & $0$ & $1$ & $\tfrac{1}{\sqrt{6}} \left(
    2|1\overline{1}2\overline{2}3\overline{3}\rangle -
    |1\overline{1}2\overline{2}4\overline{4}\rangle -
    |1\overline{1}2\overline{2}5\overline{5}\rangle\right)$
    & $(11|11) + 4(11|22) -2(12|21) + 4(11|33) -2(13|31)$ \\
    & & & & & $+ (22|22)+ 4(22|33) -2(23|32) + (33|33) -(34|43)$ \\

    \cline{5-6} & & & & $\tfrac{1}{\sqrt{6}} \left(
    2|1\overline{1}4\overline{4}5\overline{5}\rangle -
    |1\overline{1}3\overline{3}4\overline{4}\rangle -
    |1\overline{1}3\overline{3}5\overline{5}\rangle \right)$ &
    $(11|11) + 8(11|33) -4(13|31) + 2(33|33) + 4(33|44)$ \\
    & & & & &  $-3(34|43)$ \\

    \cline{5-6} & & & & cross &
    $-(23|32)$\\

    \cline{4-6} & & & $-1$ &$\tfrac{1}{\sqrt{12}} \big(
    2|1\overline{1}23\overline{4}\overline{5}\rangle
    -|1\overline{1}2\overline{3}4\overline{5}\rangle
    -|1\overline{1}2\overline{3}\overline{4}5\rangle$ &
    $(11|11) + 2(11|22) - (12|21) + 6(11|33) -3(13|31)$ \\
    & & & & $+2|1\overline{1}\overline{2}\overline{3}45\rangle
    -|1\overline{1}\overline{2}34\overline{5}\rangle
    -|1\overline{1}\overline{2}3\overline{4}5\rangle \big)$
    & $+ 3(22|33) + 3(33|44)$\\

    \cline{2-6} & $6$ & $2$ & $-1$ & $\tfrac{1}{\sqrt{6}} \big(
    2|1\overline{1}2\overline{3}45\rangle
    -|1\overline{1}234\overline{5}\rangle -
    |1\overline{1}23\overline{4}5\rangle\big)$
    & $(11|11) + 2(11|22) - (12|21) + 6(11|33) -3(13|31)$\\

    & & & & &
    $+ 3(22|33) -2(23|32) + 3(33|44)$\\
    \hline

   \end{tabular}

  } 

 \end{center}

 \caption[$V_{ee}$ matrix element expressions for C]
 {$V_{ee}$ matrix element expressions for the
 C sequence, `cross' denotes the
 off-diagonal term in the $2\times2$ matrix. See Lemma \ref{L:7.3} for notation.}
 \label{Tab:VeeMatrix2}
\end{table}

\begin{table}[htbp]
 \begin{center}

  \resizebox{\textwidth}{!}{
   \begin{tabular}{|c|c|c|c|c|l|}
     \hline & $\LL^2$&
    $\SSS^2$ &  $\hat{R}$& $\Psi$
    & \multicolumn{1}{c|}{$\langle V_{ee} \rangle$} \\
    \hline

     N & $0$ & $\tfrac{3}{4}$ & $1$ & $\tfrac{1}{\sqrt{3}} \big(
    |1\overline{1}23\overline{3}4\overline{4}\rangle +
    |1\overline{1}23\overline{3}5\overline{5}\rangle $
    & $(11|11) + 2(11|22) - (12|21) + 8(11|33) -4(13|31)$ \\
    & & & & $+|1\overline{1}24\overline{4}5\overline{5}\rangle \big)$
    &  $+ 4(22|33) -2(23|32) + 2(33|33) + 4(33|44)$ \\

    \cline{2-6} & $0$ & $\tfrac{15}{4}$ & $-1$ &
    $|1\overline{1}2\overline{2}345\rangle$
     & $(11|11) + 4(11|22) -2(12|21) + 6(11|33) -3(13|31)$  \\
    & & & & & $+ (22|22)+ 6(22|33) -3(23|32) + 3(33|44) -3(34|43)$ \\

    \cline{2-6} & $2$ & $\tfrac{3}{4}$ & $-1$ &
    $|1\overline{1}34\overline{4}5\overline{5}\rangle$  & $(11|11) +
    10(11|33) -5(13|31) + 2(33|33) + 8(33|44)$ \\
    & & & & & $ -4(34|43)$ \\

    \cline{5-6} & & & & $\tfrac{1}{\sqrt{2}}\left(
    |1\overline{1}2\overline{2}34\overline{4}\rangle +
    |1\overline{1}2\overline{2}35\overline{5}\rangle \right)$ &
    $(11|11) + 4(11|22) -2(12|21) + 6(11|33) -3(13|31)$\\
    & & & & & $+ (22|22)+ 6(22|33) -3(23|32) + (33|33) + 2(33|44)$ \\

    \cline{5-6} & & & & cross & $\sqrt{2}(23|32)$\\

    \cline{4-6} & & & $1$ & $\tfrac{1}{\sqrt{6}} \big(
    2|1\overline{1}\overline{2}3\overline{3}45\rangle -
    |1\overline{1}23\overline{3}\overline{4}5\rangle $
    & $(11|11) + 2(11|22) -(12|21) + 8(11|33) -4(13|31) $ \\
    & & & & $- |1\overline{1}23\overline{3}4\overline{5}\rangle \big)$
    &  $+ 4(22|33) + (33|33) + 5(33|44) -3(34|43)$ \\

    \cline{2-6} & $2$ & $\tfrac{15}{4}$ & $1$ &
    $|1\overline{1}23\overline{3}45\rangle$
    &  $(11|11) + 2(11|22) - (12|21) + 8(11|33) -4(13|31)$  \\
    & & & & & $+ 4(22|33) -3(23|32) + (33|33) + 5(33|44) -3(34|43)$ \\

    \cline{2-6} & $6$ & $\tfrac{3}{4}$ & $-1$ &
    $\tfrac{1}{\sqrt{6}}\big(2|1\overline{1}2\overline{2}\overline{3}45\rangle
    - |1\overline{1}2\overline{2}34\overline{5}\rangle$
    & $(11|11) + 4(11|22) -2(12|21) + 6(11|33) -3(13|31)$  \\
    & & & & $- |1\overline{1}2\overline{2}3\overline{4}5\rangle \big)$
    & $+ (22|22)+ 6(22|33) -3(23|32) + 3(33|44)$ \\

    \cline{4-6} & & & $1$ &
    $\tfrac{1}{\sqrt{6}}\big(2|1\overline{1}24\overline{4}5\overline{5}\rangle
    - |1\overline{1}23\overline{3}4\overline{4}\rangle$
    & $(11|11) + 2(11|22) -(12|21) + 8(11|33) -4(13|31)$ \\
    & & & & $-|1\overline{1}23\overline{3}5\overline{5}\rangle\big)$
    & $+ 4(22|33) -2(23|32) + 2(33|33) + 4(33|44) -3(34|43)$ \\

    \hline

    O & $0$ & $0$ & $1$ & $\tfrac{1}{\sqrt{3}}\big(
    |1\overline{1}2\overline{2}3\overline{3}4\overline{4}\rangle +
    |1\overline{1}2\overline{2}3\overline{3}5\overline{5}\rangle$ &
    $(11|11) +
    4(11|22) -2(12|21) + 8(11|33) -4(13|31)$\\
    & & & & $+
    |1\overline{1}2\overline{2}4\overline{4}5\overline{5}\rangle\big)$
    & $+
    (22|22) + 8(22|33) -4(23|32) + 2(33|33) + 4(33|44)$\\

    \cline{5-6} & & & & $|1\overline{1}3\overline{3}4\overline{4}5\overline{5}\rangle$ &
    $(11|11) + 12(11|33) -6(13|31) + 3(33|33)+ 12(33|44)$\\
    & & & & & $  -6(34|43)$ \\

    \cline{5-6} & & & & cross &
     $\sqrt{3}(23|32)$ \\

    \cline{2-6} & $2$ & $0$ & $-1$ & $\tfrac{1}{\sqrt{2}}\left(
    |1\overline{1}2\overline{3}4\overline{4}5\overline{5}\rangle-
    |1\overline{1}\overline{2}34\overline{4}5\overline{5}\rangle\right)$
    & $(11|11) + 2(11|22) -(12|21) + 10(11|33) -5(13|31)$ \\
    & & & & & $+ 5(22|33) -(23|32) + 2(33|33) + 8(33|44) -4(34|43)$ \\

    \cline{2-6} & $2$ & $2$ & $-1$ &
    $|1\overline{1}234\overline{4}5\overline{5}\rangle$
    & $(11|11) + 2(11|22) - (12|21) + 10(11|33) -5(13|31)$ \\
    & & & & & $ + 5(22|33)-3(23|32) + 2(33|33) + 8(33|44) -4(34|43)$ \\

    \cline{4-6} & & & $1$ &
    $|1\overline{1}2\overline{2}3\overline{3}45\rangle$
    & $(11|11) + 4(11|22) -2(12|21) + 8(11|33) -4(13|31)$ \\
    & & & & & $+ (22|22) + 8(22|33) -4(23|32) + (33|33) + 5(33|44)$ \\
    & & & & & $ -3(34|43)$ \\

    \cline{2-6} & $6$ & $0$ & $1$ & $\tfrac{1}{\sqrt{6}}\big(
    2|1\overline{1}2\overline{2}4\overline{4}5\overline{5}\rangle-
    |1\overline{1}2\overline{2}3\overline{3}4\overline{4}\rangle$
    & $(11|11) + 4(11|22) -2(12|21) + 8(11|33) -4(13|31)$ \\
    & & & &
    $-|1\overline{1}2\overline{2}3\overline{3}5\overline{5}\rangle
    \big)$
    & $+ (22|22) + 8(22|33) -4(23|32) + 2(33|33) + 4(33|44)$\\
    & & & & & $-3(34|43)$\\
    \hline

    F & $0$ & $\tfrac{3}{4}$ & $1$ &
    $|1\overline{1}23\overline{3}4\overline{4}5\overline{5}\rangle$
    & $(11|11) + 2(11|22) - (12|21) + 12(11|33) -6(13|31)$  \\
    & & & & & $ + 6(22|33) -3(23|32) + 3(33|33) + 12(33|44) -6(34|43)$ \\

    \cline{2-6} & $2$ & $\tfrac{3}{4}$ & $-1$ &
    $|1\overline{1}2\overline{2}34\overline{4}5\overline{5}\rangle$
    & $(11|11) + 4(11|22) -2(12|21) + 10(11|33) -5(13|31)$  \\
    & & & & & $+ (22|22) + 10(22|33) -5(23|32) + 2(33|33) + 8(33|44)$\\
    & & & & & $ -4(34|43)$ \\
    \hline

    Ne & $0$ & $0$ & $1$ &
    $|1\overline{1}2\overline{2}3\overline{3}4\overline{4}5\overline{5}\rangle$
    & $(11|11) + 4(11|22) -2(12|21)+12(11|33)-6(13|31)$\\
    & & & & & $+(22|22)+12(22|33)-6(23|32)+3(33|33) + 12(33|44)$ \\
    & & & & & $-6(34|43)$ \\
    \hline

   \end{tabular}
  }  

 \end{center}

 \caption[$V_{ee}$ matrix element expressions for N-Ne]
 {$V_{ee}$ matrix element expressions for the
 N-Ne sequences, `cross' denotes the
 off-diagonal term in the $2\times2$ matrix. See Lemma \ref{L:7.3} for notation.}
 \label{Tab:VeeMatrix3}
\end{table}


\subsection{Explicit interaction matrix}
In order to obtain explicit values, we
finally need to substitute the explicit PT orbitals (\ref{PTorbitals}), (\ref{porbitals}), and
evaluate the ensuing Coulomb and exchange integrals.
We do this via a four-step procedure: reduce the original integrals over $\R^6$ to
integrals over $\R^3$ via Fourier transform calculus; explicitly determine the Fourier transforms
of pointwise products of the above orbitals; reduce to 1D integrals with the help of spherical
polar coordinates in Fourier space; evaluate the remaining 1D integrals, which turn out to have
rational integrands.

The Fourier transform of a function $f\in L^1(\R^n)$ will be
denoted $\hat{f}$; we find it convenient to use the definition
\be \label{Fourier}
        \hat{f}(k):= \int_{\R^n} f(x) e^{-ik\cdot x} dx
\ee
which does not contain any normalization constants.

\begin{lemma}\label{L:CoulombIntegrals}
For one-electron orbitals $\psi_\alpha \in L^2(\R^3)\cap
L^\infty(\R^3)$, with $\hat\psi_\alpha \in L^2(\R^3) \cap
L^\infty(\R^3)$ and $\alpha \in \{i,j,k,\ell\}$, let
$f(x):=\psi_i(x)\psi_j^*(x)$ and $g(x):=\psi_k^*(x)\psi_\ell(x)$.
Then
\begin{align}
   (\psi_i \psi_j | \psi_k \psi_\ell)
   &= \int_{\R^6} dx_1 dx_2 \psi_i^*(x_1)\psi_j(x_1)
   \frac{1}{|x_1-x_2|} \psi_k^*(x_2)\psi_\ell(x_2) \notag \\
   &= \frac{1}{2\pi^2}\int_{\R^3}dk \frac{1}{|k|^2}
   (\widehat{f})^*(k)\widehat{g}(k). \label{ijklFT}
\end{align}
\end{lemma}
\noindent
Note that this shows that exchange integrals $(\psi_i\psi_j|\psi_j\psi_i)$ are positive.
\\[2mm]
\noindent\textbf{Proof} Since the $\psi_\alpha$ are in
$L^2(\R^3)\cap L^\infty(\R^3)$, their products $f$ and $g$ are in
$L^1(\R^3)\cap L^\infty(\R^3)$ and hence their Fourier transforms
are well defined.  Considering the integral
$I(\lambda)=\int_{\R^3}\int_{\R^3} dx dy
\frac{e^{-\lambda|x-y|}}{|x-y|} f^*(x)g(y)$, $\lambda>0$.  It is
easy to show that $\widehat{\frac{e^{-\lambda|x|}}{|x|}}
=\frac{4\pi}{\lambda^2+|k|^2}$.  Since $f,g \in L^1(\R^3)\cap
L^\infty(\R^3)$ it follows that $|\frac{e^{-\lambda|x-y|}}{|x-y|}
f^*(x)g(y)| \in L^1(\R^6)$ and so by dominated convergence
\be
    I(\lambda) \to \int_{\R^3}\int_{\R^3} dx dy \frac{1}{|x-y|}
    f^*(x)g(y) \quad (\lambda \to 0). \label{DCT1}
\ee

Setting $h=e^{-\lambda|x|}/|x|$ we have $I(\lambda)= \int_{\R^3}dy
\big(f \ast h\big)(y)g(y)dy$ and $\widehat{f\ast h}=\hat{f}\hat{h}
\in L^1(\R^3)$ since $\hat{f}\in L^1(\R^3)$ and $\hat{h}\in
L^\infty(\R^3)$.  By Plancherel's theorem we have
$I(\lambda)=\frac{1}{(2\pi)^3}\int_{\R^3} dk \frac{4\pi}{\lambda^2 +
|k|^2}(\hat{f})^*(k)\hat{g}(k)$, and again by dominated convergence and using $f,g \in
L^1(\R^3) \cap L^\infty(\R^3)$,
\be
    I(\lambda) \to \frac{1}{2\pi^2}\int_{\R^3}dk \frac{1}{|k|^2}
    (\widehat{f})^*(k)\widehat{g}(k) \quad (\lambda \to 0).
    \label{DCT2}
\ee
Combining (\ref{DCT1}) and (\ref{DCT2}) gives the result.
\qed \\

Next we calculate the Fourier transforms of pointwise products of
the hydrogen orbitals (\ref{PTorbitals}), (\ref{porbitals}). Here and below, by expressions such as
$\widehat{|x|e^{-\lambda |x|}}(k)$ we mean the Fourier transform
$\widehat{f}(k)$ of the function $f(x)=|x|e^{-\lambda|x|}$.
\begin{lemma} \label{L:FTs}
With the Fourier transform as defined in (\ref{Fourier}), and $\lambda>0$,
\begin{center}
\begin{tabular}{|c|c|}
\hline Function & Fourier transform \\
\hline

$e^{-\lambda |x|}$ & $\frac{8 \lambda \pi}{(\lambda^2 +|k|^2)^2}$ \\

$|x|e^{-\lambda |x|}$ & $\frac{32 \lambda^2 \pi}{(\lambda^2 + |k|^2)^3}
-\frac{8\pi}{(\lambda^2 + |k|^2)^2}$ \\

$|x|^2e^{-\lambda |x|}$ & $\frac{192\lambda^3 \pi}{(\lambda^2+|k|^2)^4} - \frac{96\lambda
\pi}{(\lambda^2+|k|^2)^3}$ \\

$x_j e^{-\lambda |x|}$ & $-\frac{32 i \lambda \pi
k_j}{(\lambda^2+|k|^2)^3}$\\

$x_j^2 e^{-\lambda |x|}$ & $\frac{32\lambda \pi}{(\lambda^2+|k|^2)^3}
- \frac{192  \lambda \pi k_j^2}{(\lambda^2+|k|^2)^4}$\\

$x_{\ell}x_j e^{-\lambda |x|}$ \scriptsize{$(j \neq \ell)$} & $ - \frac{192  \lambda \pi
k_j k_{\ell}}{(\lambda^2+|k|^2)^4}$\\

$|x|x_je^{-\lambda |x|}$ & $\frac{32 i \pi k_j}{(\lambda^2+|k|^2)^3} - \frac{192 i
\lambda^2 \pi k_j}{(\lambda^2+|k|^2)^4}$\\
\hline
\end{tabular}
\end{center}

\end{lemma}
\noindent \textbf{Proof} Let $f(x):=e^{-\lambda |x|}$. We have
   $\widehat{f}(k)=8\pi\lambda/(\lambda^2+|k|^2)^2$,
which is easy to verify by
direct calculation (convert to spherical polar coordinates and
integrate). All remaining Fourier transforms can be expressed in terms of derivatives of $\widehat{f}$, as follows.
Using $|x|e^{-\lambda |x|}=-\frac{d}{d\lambda}e^{-\lambda |x|}$, $|x|^2e^{-\lambda |x|}=\frac{d^2}{d\lambda^2}e^{-\lambda |x|}$,
and noting that differentiation with respect to $\lambda$ commutes
with the Fourier transform gives
$$
    \widehat{|\cdot|e^{-\lambda |\, \cdot \, |}}(k)=-\frac{\partial}{\partial \lambda}\widehat{f}(k), \;\;
    \widehat{|\cdot|^2e^{-\lambda |\cdot|}}(k) =\frac{d^2}{d\lambda^2}\widehat{f}(k).
$$
For the next three Fourier transforms, we recall the well known
differentiation identities for Fourier transforms:
$$
    \widehat{x_j f}(k) =i \frac{\partial}{\partial k_j}\widehat{f}(k)
    \text{ and } \widehat{x_{\ell}x_j f}(k) =-
    \frac{\partial^2}{\partial k_{\ell} \partial k_j}\widehat{f}(k).
$$
Consequently
$$
    \widehat{(x_j e^{-\lambda|x|})}(k) \!=\! i\frac{\partial}{\partial k_j}\widehat{f}(k), \;
    \widehat{x_j x_\ell e^{-\lambda|x|}}(k) \!=\! - \frac{\partial^2}{\partial k_j \partial k_\ell}\widehat{f}(k).
$$
The final case needed is
$$
    \widehat{(|x|x_j e^{-\lambda|x|})}(k) = - \frac{d}{d\lambda} \Bigl(i\frac{\partial}{\partial k_j}\widehat{f}(k)\Bigr).
$$
Working out the above derivatives of $\widehat{f}$ explicitly is straightforward, yielding the formulae given in the lemma. \qed
\begin{lemma} \label{L:OrbitalFTs}
The Fourier transforms of pointwise products of the hydrogen orbitals
(\ref{PTorbitals}), (\ref{porbitals}) are as given in the following table.  In
all cases $j, \ell =1,2,3$, $j \neq \ell$.
\begin{center}
 \begin{tabular}{|c|c|}
   \hline Function & Fourier Transform \\
   \hline $\phi_{1s}\phi_{1s}$ & $\frac{16Z^4}{(4Z^2+|k|^2)^2}$ \\
   $\phi_{2s}\phi_{2s}$ & $\frac{2Z^4}{(Z^2+|k|^2)^2}-
   \frac{7Z^6}{(Z^2+|k|^2)^3} +
   \frac{6Z^8}{(Z^2+|k|^2)^4}$\\
   $\phi_{1s}\phi_{2s}$ & $\frac{4\sqrt{2}Z^4}{((\frac{3}{2}Z)^2+|k|^2)^2}
   -\frac{9\sqrt{2}Z^6}{((\frac{3}{2}Z)^2+|k|^2)^3}$ \\
   $\phi_{2p_j}\phi_{2p_j}$ &
$\frac{Z^6}{(Z^2+|k|^2)^3}-\frac{6Z^6k_j^2}{(Z^2+|k|^2)^4}$ \\
   $\phi_{1s}\phi_{2p_j}$& $-\frac{6\sqrt{2}iZ^5 k_j}{((\frac{3}{2}Z)^2 +|k|^2)^3}$
\\
   $\phi_{2s}\phi_{2p_j}$& $\frac{6Z^7 i
k_j}{(Z^2+|k|^2)^4}-\frac{3Z^5ik_j}{(Z^2+|k|^2)^3}$ \\
   $\phi_{2p_j}\phi_{2p_\ell}$& $-\frac{6k_jk_{\ell}Z^6}{(Z^2+|k|^2)^4}$ \\
   \hline
  \end{tabular}
\end{center}
\end{lemma}
\noindent \textbf{Proof} This is simply an application of the
results of Lemma \ref{L:FTs}. \qed
\\[2mm]
Finally we use these Fourier transforms, along with the
reformulation of the Coulomb and exchange integrals from Lemma
\ref{L:CoulombIntegrals}, to derive the explicit values of these
integrals.

\begin{lemma}\label{L:VeeElements}
Using the abbreviated notation
$1=\phi_{1s}$, $2=\phi_{2s}$, $3=\phi_{2p_3}$, $4=\phi_{2p_1}$, $5=\phi_{2p_1}$, the Coulomb and
exchange integrals (\ref{ijklIntegral}) occuring in Tables \ref{Tab:VeeMatrix1}--\ref{Tab:VeeMatrix3}
with the PT orbitals (\ref{PTorbitals}), (\ref{porbitals}) are given by
\vspace*{-3mm}
 \begin{center}
 \resizebox{\textwidth}{!}{
 \begin{tabular}{|c|c|c|c|c|c|c|c|c|c|c|}
    \hline
    $(11|11)$ & $(11|22)$ & $(12|21)$ & $(22|22)$
  & $(11|33)$ & $(13|31)$ & $(22|33)$ & $(23|32)$
  & $(33|33)$ & $(33|44)$ & $(34|43)$ \\
    \hline
    $\frac{5}{8}Z$ & $\frac{17}{81}Z$ & $\frac{16}{729}Z$ & $\frac{77}{512}Z$
  & $\frac{59}{243}Z$ &  $\frac{112}{6561}Z$ & $\frac{83}{512}Z$ & $\frac{15}{512}Z$
  & $\frac{501}{2560} Z$ & $\frac{447}{2560} Z$ & $\frac{27}{2560} Z$ \\
    \hline
  \end{tabular}
  } 
  \end{center}
\end{lemma}
\noindent \textbf{Proof} We insert the Fourier transforms from Lemma
\ref{L:OrbitalFTs} into (\ref{ijklFT}), change to spherical polar coordinates and
integrate. The angular integrals are elementary and the final radial
integrals, which on account of Lemmas \ref{L:OrbitalFTs} and \ref{L:CoulombIntegrals} have
rational integrands, may be evaluated with Maple; for a truly pen and paper
method, one can use complex contour integration. \qed

Note the insteresting multiscale effect that the exchange integrals are much smaller then the Coulomb integrals,
by about one order of magnitude. Nevertheless, as we will see later, the exchange terms play an important role
in energy level splitting.

The table in the above lemma together with Tables \ref{Tab:VeeMatrix1}--\ref{Tab:VeeMatrix3} completes the task of
evaluating the matrix $PV_{ee}P$ on $V_0(N)$.

\subsection{The matrix $PHP$} \label{Sec:ExplicitPHP}

The remaining part $PH_0P$ of the Hamiltonian $PHP$ is trivial
to determine, because the space $V_0(N)$ is an eigenspace
of $H_0$, with eigenvalue given in Lemma \ref{L:meEfns}:
\be \label{H0PT}
     H_0 = PH_0P = Z^2\Bigl(-1-\mbox{$\frac{N-2}{8}$}\Bigr)I \;\;\mbox{on }V_0(N).
\ee
By inspection of (\ref{H0PT}), Lemma \ref{L:VeeElements}, and Tables \ref{Tab:VeeMatrix1}--\ref{Tab:VeeMatrix3}
we obtain an interesting corollary.
\begin{corollary} The matrix of the PT Hamiltonian $PHP$ with respect to the basis in Tables
\ref{Tab:VeeMatrix1}--\ref{Tab:VeeMatrix3}
with the orbitals (\ref{PTorbitals}), (\ref{porbitals}) is a rational matrix.
\end{corollary}
\section{Atomic energy levels and eigenstates} \label{Sec:AtSpec}
The spectral decomposition of the PT Hamiltonians $PHP$ is almost immediate from the block form
derived in the previous section , the only remaining task being the
diagonalization of the $2\times 2$ blocks, which may be done explicitly:
For (orthonormal) wavefunctions $\Psi_1$, $\Psi_2$
and $E_i:=\langle \Psi_i |  H  | \Psi_i\rangle$, the eigenvalues are given by
\be \label{degevals}
    \lambda_{\pm} = \frac{E_1+E_2}{2} \pm
    \sqrt{\big(\tfrac{E_1-E_2}{2}\big)^2 +
    |\langle \Psi_1 |V_{ee} |\Psi_2\rangle|^2}
\ee
with corresponding normalized eigenstates
\be \label{degestates}
    \Psi_\pm = \frac{1}{\sqrt{1+c_\pm^2}}\Bigl( \Psi_1 + c_\pm \Psi_2 \Bigr), \;\;\;
    c_\pm = \frac{ \frac{E_2-E_1}{2} \pm  \sqrt{\big(\tfrac{E_1-E_2}{2}\big)^2 +
    |\langle \Psi_1 |V_{ee} |\Psi_2\rangle|^2     }          }{\langle \Psi_1 |V_{ee} |\Psi_2\rangle}.
\ee
These formulae together with eq. (\ref{H0PT}), Tables \ref{Tab:VeeMatrix1}--\ref{Tab:VeeMatrix3} and
Lemma \ref{L:VeeElements} immediately yield:

\begin{theorem} \label{mainresult} For $N=3,...,10$, and $Z>0$, the
energy levels of the PT Model (\ref{PT}), (\ref{H0H'}) are as given in
Tables \ref{Tab:PTEigenspaces1}, \ref{Tab:PTEigenspaces2}. Each eigenspace has
the minimal dimension $(2L+1)(2S+1)$ possible for its spin and angular momentum quantum numbers $L$ and $S$ (see Lemma \ref{symlem}),
and the up to normalization unique corresponding
eigenstate with zero $L_3$ and maximal $S_3$ is as shown in the Tables. Moreover the levels and eigenstates in the Tables provide the
leading order asymptotic terms of the true Schr\"odinger levels as $Z\to\infty$, in the sense described in Theorem \ref{T:isolimit}.
\end{theorem}

\noindent
Note that the ordering of the PT levels is independent of $Z$, since the spectral gaps are linear in $Z$.

\begin{table}[htbp]
\hspace{5mm}
\rotatebox{90}{ %
\resizebox{0.9\textheight}{!}{
  \begin{tabular}{|c|c|c|l|c|r|r|}
    \hline & Symm. & $\Psi$ &
    \multicolumn{1}{|c|}{$E$} &
    \multicolumn{1}{|c|}{$c$} & \multicolumn{1}{|c|}{$E$ (num.)} &
    \multicolumn{1}{|c|}{$c$ (num.)} \\
    \hline
    Li & $^2S$    & $\Psi_1$ & $-\tfrac{9}{8}Z^2+\tfrac{5965}{5832}Z$ & &-7.0566 & \\
    & $^2P^\circ$ & $\Psi_2$ & $-\tfrac{9}{8}Z^2+\tfrac{57397}{52488}Z$ & &-6.8444 &\\ \hline
    Be & $^1S$    & $\frac{1}{\sqrt{1+c^2}}(\Psi_1 + c \Psi_2)$
    & $-\frac{5}{4}Z^2 +
    \frac{1}{1679616}(2813231-5\sqrt{1509308377})Z$
    & $-\frac{1}{59049}(2\sqrt{1509308377}-6981)\sqrt{3}$
    &-13.7629 & -0.2311\\
    & $^3P^\circ$ & $\Psi_4$ & $-\frac{5}{4}Z^2 + \frac{1363969}{839808}Z$ & &-13.5034 &\\
    & $^1P^\circ$ & $\Psi_3$ & $-\frac{5}{4}Z^2 + \frac{2826353}{1679616}Z$& &-13.2690 &\\
    & $^3P$       & $\Psi_5$ & $-\frac{5}{4}Z^2 + \frac{1449605}{839808}Z$ & &-13.0955 &\\
    & $^1D$       & $\Psi_6$ & $-\frac{5}{4}Z^2 + \frac{14673197}{8398080}Z$ & &-13.0112 &\\
    & $^1S$       & $\frac{1}{\sqrt{1+c^2}}(\Psi_1 + c \Psi_2)$
    & $-\frac{5}{4}Z^2 +
    \frac{1}{1679616}(2813231+5\sqrt{1509308377})Z$
    & $\frac{1}{59049}(2\sqrt{1509308377}+6981)\sqrt{3}$
    &-12.8377 & 4.3271\\
    \hline
    B & $^2P^\circ$ & $\frac{1}{\sqrt{1+c^2}}(\Psi_3 + c \Psi_4)$
    & $-\frac{11}{8}Z^2 +
    \frac{1}{6718464}(16493659-\sqrt{733174301809})Z$
    & $-\frac{1}{393660}(\sqrt{733174301809}-809747)\sqrt{2}$
    &-22.7374 & -0.1671\\
    & $^4P$ & $\Psi_6$ & $-\frac{11}{8}Z^2 + \frac{2006759}{839808}Z$ & &-22.4273 &\\
    & $^2D$ & $\Psi_7$ & $-\frac{11}{8}Z^2 + \frac{40981549}{16796160}Z$& &-22.1753 &\\
    & $^2S$ & $\Psi_1$ & $-\frac{11}{8}Z^2+ \frac{4151299}{1679616}Z$& &-22.0171 &\\
    & $^2P$ & $\Psi_5$ & $-\frac{11}{8}Z^2 + \frac{8322281}{3359232}Z$& &-21.9878 &\\
    & $^4S^\circ$ & $\Psi_2$ & $-\frac{11}{8}Z^2 + \frac{706213}{279936}Z$ & &-21.7612 &\\
    & $^2D^\circ$ & $\Psi_8$ & $-\frac{11}{8}Z^2 + \frac{14301407}{5598720}Z$ & &-21.6030 &\\
    & $^2P^\circ$ & $\frac{1}{\sqrt{1+c^2}}(\Psi_3 + c \Psi_4)$
    & $-\frac{11}{8}Z^2 +
    \frac{1}{6718464}(16493659+\sqrt{733174301809})Z$
    & $\frac{1}{393660}(\sqrt{733174301809}+809747)\sqrt{2}$
    &-21.4629 & 5.9851\\
    \hline
    C & $^3P$ & $\frac{1}{\sqrt{1+c^2}}(\Psi_6+c\Psi_7)$
    & $-\frac{3}{2}Z^2 + \big( \frac{3806107}{1119744}
    - \frac{1}{3359232}\sqrt{221876564389} \big)Z$
    & $-\frac{1}{98415}(\sqrt{221876564389}-460642)$
    &-34.4468 & -0.1056\\
    & $^1D$ & $\frac{1}{\sqrt{1+c^2}}(\Psi_9+c\Psi_{10})$
    & $-\frac{3}{2}Z^2 + \big( \frac{19148633}{5598720}
    -\frac{1}{3359232}\sqrt{221876564389} \big)Z$
    & $\frac{1}{98415}(\sqrt{221876564389}-460642)$
    &-34.3202 & 0.1056\\
    & $^5S^\circ$ & $\Psi_4$ & $-\frac{3}{2}Z^2 +
    \frac{464555}{139968}Z $
    & &-34.0859 &\\
    & $^1S$ & $\frac{1}{\sqrt{1+c^2}}(\Psi_1+c\Psi_2)$
    & $-\frac{3}{2}Z^2 + \big( \frac{966289}{279936}
    - \frac{1}{1679616}\sqrt{62733275266}\big)Z$
    & $-\frac{1}{98415}(\sqrt{62733275266}-230321)$
    &-34.1838 & -0.2047\\
    & $^3D^\circ$ & $\Psi_{12}$ &
    $-\frac{3}{2}Z^2 + \frac{4730843}{1399680}Z$ & &-33.7203 &\\
    & $^3P^\circ$ & $\Psi_8$ &
    $-\frac{3}{2}Z^2 + \frac{1904147}{559872}Z$ & &-33.5938 &\\
    & $^1D^\circ$ & $\Psi_{11}$ &
    $-\frac{3}{2}Z^2 + \frac{9625711}{2799360}Z$ & &-33.3688 &\\
    & $^3S^\circ$ & $\Psi_3$ &
    $-\frac{3}{2}Z^2 + \frac{961915}{279936}Z$ & &-33.3828 &\\
    & $^1P^\circ$ & $\Psi_5$
    & $-\frac{3}{2}Z^2 + \frac{242119}{69984}Z$ & &-33.2422 &\\
    & $^3P$ & $\frac{1}{\sqrt{1+c^2}}(\Psi_6+c\Psi_7)$
    & $-\frac{3}{2}Z^2 + \big( \frac{3806107}{1119744}
    + \frac{1}{3359232}\sqrt{221876564389} \big)Z$
    & $\frac{1}{98415}(\sqrt{221876564389}+460642)$
    &-32.7641 & 9.4668\\
    & $^1D$ & $\frac{1}{\sqrt{1+c^2}}(\Psi_9+c\Psi_{10})$
    & $-\frac{3}{2}Z^2 + \big( \frac{19148633}{5598720}
    +\frac{1}{3359232}\sqrt{221876564389} \big)Z$
    & $\frac{1}{98415}(-\sqrt{221876564389}-460642)$
    &-32.6376 & -9.4668\\
    & $^1S$ & $\frac{1}{\sqrt{1+c^2}}(\Psi_1+c\Psi_2)$
    & $-\frac{3}{2}Z^2 + \big( \frac{966289}{279936}
    + \frac{1}{1679616}\sqrt{62733275266}\big)Z$
    & $\frac{1}{98415}(\sqrt{62733275266}+230321)$
    &-32.3943 & 4.8853\\
    \hline
  \end{tabular}
} 
} 
 \caption{Asymptotic Schr\"odinger energy levels and eigenstates (= exact PT levels and states) for the Li-C sequences. The
numerical values are for $Z\!=\!N$. $\Psi$ is the up to
normalization unique PT eigenstate with $L_3\!=\!0$ and maximal $S_3$. The $\Psi_i$
are as in Tables \ref{Tab:VeeMatrix1}-\ref{Tab:VeeMatrix2}, labelled by order of appearance, with
$1,2,3,4,5,$ $\overline{1},\overline{2},\overline{3},\overline{4},\overline{5}$ as in Theorem \ref{GStheorem}.}
 \label{Tab:PTEigenspaces1}
\end{table}

\begin{table}[htbp]
\hspace{25mm}
\rotatebox{90}{ %
\resizebox{0.9\textheight}{!}{
  \begin{tabular}{|c|c|c|l|c|r|r|}
    \hline & Symm. & $\Psi$ &
    \multicolumn{1}{|c|}{$E$} &
    \multicolumn{1}{|c|}{$c$} & \multicolumn{1}{|c|}{$E$ (num.)} &
    \multicolumn{1}{|c|}{$c$ (num.)} \\
    \hline
    N & $^4S^\circ$ & $\Psi_2$ & $-\frac{13}{8}Z^2 + \frac{2437421}{559872}Z$ & & -49.1503 &  \\
    & $^2D^\circ$ & $\Psi_7$ & $-\frac{13}{8}Z^2 + \frac{24551357}{5598720}Z$ & &-48.9288 & \\
    & $^2P^\circ$ & $\frac{1}{\sqrt{1+c^2}}(\Psi_4+c\Psi_3)$
    & $-\frac{13}{8}Z^2 + \frac{1}{6718464}(30636167 -
    \sqrt{1144203315841})Z$
    & $-\frac{1}{393660}(\sqrt{1144203315841}-1032821)\sqrt{2}$
    &-48.8195 & -0.1324 \\
    & $^4P$ & $\Psi_6$ & $-\frac{13}{8}Z^2 + \frac{7549145}{1679616}Z$ & &-48.1630 & \\
    & $^2D$ & $\Psi_8$ & $-\frac{13}{8}Z^2 + \frac{76337819}{16796160}Z$& &-47.8103 & \\
    & $^2S$ & $\Psi_1$ & $-\frac{13}{8}Z^2 + \frac{3843463}{839808}Z$ & &-47.5888 &\\
    & $^2P$ & $\Psi_5$ & $-\frac{13}{8}Z^2 + \frac{15393535}{3359232}Z$& &-47.5478 &\\
    & $^2P^\circ$ & $\frac{1}{\sqrt{1+c^2}}(\Psi_4+c\Psi_3)$
    & $-\frac{13}{8}Z^2 + \frac{1}{6718464}(30636167 +
    \sqrt{1144203315841})Z$
    & $\frac{1}{393660}(\sqrt{1144203315841}+1032821)\sqrt{2}$
    &-46.5905 & 7.5532\\
    \hline
    O & $^3P$ & $\Psi_5$ & $-\frac{7}{4}Z^2 + \frac{4754911}{839808}Z$ & &-66.7048 &\\
    & $^1D$ & $\Psi_6$ & $-\frac{7}{4}Z^2 + \frac{47726257}{8398080}Z$ & &-66.5360 &\\
    & $^1S$ & $\frac{1}{\sqrt{1+c^2}}(\Psi_1+c\Psi_2)$
    & $-\frac{7}{4}Z^2 +
    \frac{1}{1679616}(9884485-\sqrt{89111336179})Z$
    & $-\frac{1}{295245}(2\sqrt{89111336179}-572179)\sqrt{3}$
    &-66.3421 & -0.1458\\
    & $^3P^\circ$ & $\Psi_4$ & $-\frac{7}{4}Z^2 + \frac{1224899}{209952}Z$ & &-65.3265 &\\
    & $^1P^\circ$ & $\Psi_3$ & $-\frac{7}{4}Z^2 + \frac{9897607}{1679616}Z$ & &-64.8578 &\\
    & $^1S$ & $\frac{1}{\sqrt{1+c^2}}(\Psi_1+c\Psi_2)$
    & $-\frac{7}{4}Z^2 +
    \frac{1}{1679616}(9884485+\sqrt{89111336179})Z$
    & $\frac{1}{295245}(2\sqrt{89111336179}+572179)\sqrt{3}$
    &-63.4984 & 6.8592\\
    \hline
    F & $^2P^\circ$ & $\Psi_2$ & $-\frac{15}{8}Z^2 + \frac{11982943}{1679616}Z$ & &-87.6660 &\\
    & $^2S$ & $\Psi_1$ & $-\frac{15}{8}Z^2 +  \frac{4108267}{559872}Z$ & &-85.8342 &\\
    \hline
    Ne & $^1S$ & $\Psi_1$ & $-2Z^2 + \frac{2455271}{279936}Z$ & &-112.2917 &\\
    \hline
  \end{tabular}
} 
} 
 \caption{Asymptotic Schr\"odinger energy levels and eigenstates (= exact PT levels and states) for the N-Ne sequences.
The numerical values are for $Z=N$. $\Psi$ is the up to
normalization unique PT eigenstate with $L_3=0$ and maximal $S_3$.
The $\Psi_i$ are as in Table \ref{Tab:VeeMatrix3}, labelled by order of appearance, with
$1,2,3,4,5,\overline{1},\overline{2},\overline{3},\overline{4},\overline{5}$ as in Theorem \ref{GStheorem}.}
 \label{Tab:PTEigenspaces2}
\end{table}

\section{Comparison with experiment and methods in the physics and chemistry literature}\label{Sec:Comparison}

The analytical eigenvalues and eigenstates derived in the isoelectronic limit
provide a wealth of information on various quantitites of physical and chemical interest,
and yield a number of insights into the inner working mechanisms of the many-electron Schr\"odinger equation which
are not readily available from numerical simulations.

We will discuss, in turn, the obtained $L$ and $S$ values, ground state dimensions, ground state energies,
and spectral orderings.

\subsection{$L$ and $S$ values and the notion of `group' in the periodic table}
The ground states themselves are not accessible from experiment, but
their spin and angular momentum quantum numbers are. As already mentioned, the theoretical values agree with the experimental
values in every case (see Table \ref{Tab:H0Dim}), not just for large $Z$ but all the way down to neutral atoms
($Z=N$), capturing the nontrivial dependence on the number of electrons.

An important theoretical feature of $L$ and $S$ values as compared to the more familiar semi-empirical concept
of ``hydrogen orbital configurations'' is that regardless of the approximations made to predict them in practice,
they remain well defined in the full Schr\"odinger equation. See Section \ref{Sec:Basic}.
It would therefore be of value to base quantum mechanical explanations of the periodic table on numbers
such as these. In this context we note that $L$ and $S$ values suffice to explain quantum mechanically
a large part of the notion of ``group'' in the periodic table.
Only five different $(L,S)$ pairs occur mathematically for the first 10 atoms, and experimentally for the first 20. Now
these correspond precisely to group 1 (alkali metals),
the union of groups 2 and 8 (alkaline earth metals and noble gases),
the union of groups 3 and 7 (group 3 metals and halogens), the union of groups 4 and 6 (Carbon group and Oxygen group),
and group 5 (Nitrogen group). See the table below.

\begin{figure}[ht] \label{F:periodictable}
  \begin{center}
   \resizebox{11cm}{!}{
    \includegraphics{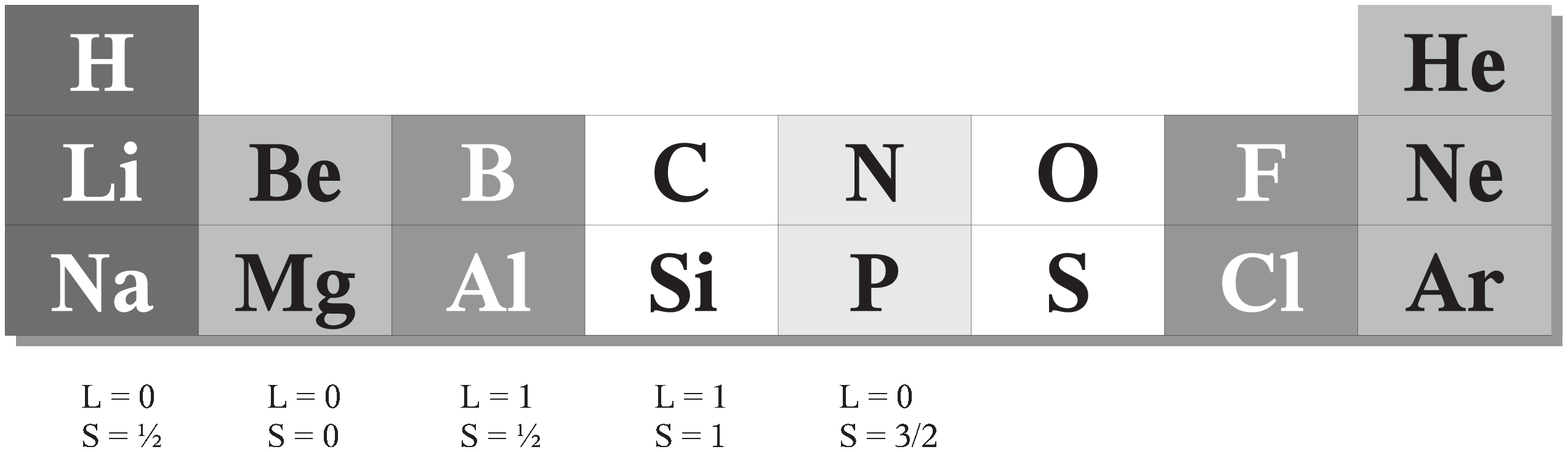}
   }
  \end{center}
\end{figure}
\noindent
Moreover, taking into account the gradients of $L$ or $S$ with respect to
atomic number $N$ would separate the group 3 metals from the halogens, and the Carbon group from the Oxygen group.
Note that $L$ and $S$ gradients are mathematically analogous to ionization energies, which are gradients of energy with respect to $N$.

\subsection{Ground state dimensions} These dimensions are shown in Table \ref{Tab:H0Dim}. They are interesting
as they are a measure of the `flexibility' within the ground state, in that they specify the number of degrees of freedom which can be varied without
affecting the energy of the state. This flexibility appears to be curiously unexplored in the literature, perhaps in part due to it not being
clearly captured by the semi-empirical Bohr-Slater picture, the Hartree-Fock approximation, or Kohn-Sham density
functional theory.

On a qualitative level, we expect that an atom with a high-dimensional ground state will form a wider range of molecules than an atom
with a similar number of valence electrons but with a lower dimensional ground state. This should be true both in terms of
molecular geometry (e.g. linear, bent, triangular, tetrahedral) as well as in terms of which atoms it will stably bond with. We plan
to develop this idea in a more mathematical way in a future publication.
\subsection{Ground state energies} \label{Sec:GSE}
The asymptotic ground state energies, despite being theoretically justified only for strongly positive ions
(see Section \ref{Sec:Pert}), still capture around $90\%$ of
the experimental \cite{NIST} energies of neutral atoms. See the following table.
\\[1mm]
  \resizebox{\textwidth}{!} 
  {
  \begin{tabular}{|l|c|c|c|c|c|c|c|c|}
    \hline
    Atom & Li & Be & B   & C   & N   & O   & F   & Ne \\
    \hline
    $E_{PT}$  & -7.0566 & -13.7629 & -22.7374 & -34.4468 & -49.1503 & -66.7048 & -87.6660 & -112.2917 \\
    $E_{exp}$ & -7.4779 & -14.6684 & -24.6581 & -37.8558 & -54.6117 & -75.1080 & -99.8060 & -129.0500 \\
    Error & 5.6\% & 6.2\% & 7.8\% & 9.0\% & 10.0\% & 11.2\% &
    12.2\% & 13.0\% \\
    \hline
  \end{tabular}
  } 
%
%
%

\subsection{Spectral orderings and Hund's rule} \label{Sec:Hund}
The spectral orderings of the asymptotic levels are in spectacular
agreement with the experimental data \cite{Huheey93,NIST}, even for neutral atoms. For the
purpose of these comparisons we consider only the experimental
states attributed to configurations containing only orbitals with $n
\leq 2$. The results differ only by the interchange of two higher levels
in Beryllium (${}^1D$ and ${}^3P$) and Carbon (${}^1D^o$ and ${}^3S^o$).
\\[2mm]
A key virtue of our exact eigenstates is that they allow to trace the spectral
gaps to the size of individual Coulomb and exchange integrals.

As an example of a $2s$--$2p$ spectral gap, consider the $^2S$ ground
state and $^2P$ first excited state of Lithium. Table
\ref{Tab:VeeMatrix2} shows that the gap is given by
the difference in interaction of the $2p$ and $2s$ orbitals with the
$1s$ shell, $[2(11|33)-(13|31)] - [2(11|22)-(12|21)]$.

As an example of energy level splitting between two states with an equal number of
$1s$, $2s$ and $2p$ orbitals, consider the $^4S^o$ ground state and $^2D^o$ first
excited state of Nitrogen. A look at Table \ref{Tab:VeeMatrix3}
reveals that the energy difference consists only of
the exchange term $-3(34|43)$, which is present in the ground state
due to the parallel spins of the three $p$-orbitals, but absent in
the excited state.
\\[2mm]
In a large majority of cases, the theoretical orderings also agree with Hund's rules.
In fact, many of Hund's rules are rigorous theorems in first order perturbation theory and
related models, and rely only on the structure of the symbolic
matrices in Tables \ref{Tab:VeeMatrix1}--\ref{Tab:VeeMatrix3}, not their numerical values. This will be discussed elsewhere.

Let us also describe a
\\[2mm]
{\it Counterexample to Hund's rules}. Consider the higher Carbon $1s^22s2p^3$ states. Hund's rules would
order their energies, regardless of the choice of nuclear charge $Z$, as
$$
   E_{^5S^o}<E_{^3D^o}<E_{^3P^o}<E_{^3S^o}<E_{^1D^o}<E_{^1P^o}.
$$
For large $Z$ this agrees with the PT and experimental orderings. (That the latter two agree with
each other follows from Theorem \ref{T:isolimit}.)

But experimentally, at $Z=20$ the
$^1D^o$ singlet and the $^3S^o$ triplet are observed to cross, see \cite{NIST} and Figure \ref{F:Splitting}.
(This crossing is beautifully confirmed by theoretical calculations based on a simple CI model designed with
the help of our asymptotic findings here, see \cite{FG09}.)
In particular, in the neutral atom, $Z=N$, the experimental ordering is
$$
   E_{^5S^o}<E_{^3D^o}<E_{^3P^o}<E_{^1D^o}<E_{^3S^o}<E_{^1P^o}.
$$
This is an important example because it shows that it is of
value to investigate which of Hund's rules can be justified quantum
mechanically and which ones can not. In this particular case, a
closer look shows that there should be no universal ordering,
neither one way nor the other. The energy difference as read off
from Table \ref{Tab:VeeMatrix2} consists of a $2s$--$2p$ positive
exchange term and a $2p$--$2p$ negative exchange term
\be \label{truediff}
   E_{^3S^o} - E_{^1D^o} = (24|42) - 3(34|43),
\ee
and so could have either sign, depending on the orbitals.

Note that this interesting effect is missed when the states under investigation are modelled
by their aufbau principle configurations. By Hund's rules, these are
$|1\overline{1}23\overline{4}\overline{5}\rangle$ for the singlet
and $|1\overline{1}234\overline{5}\rangle$ for the triplet. A simple
calculation shows that the energy difference is then
$E_{^3S^o} - E_{^1D^o} = - (24|42) < 0$, which is very far from the correct difference (\ref{truediff})
and incorrectly predicts a universal ordering.

\subsection{Spectral gaps}
\begin{figure}[ht]
  \begin{center}
   \resizebox{8cm}{!}{
    \includegraphics{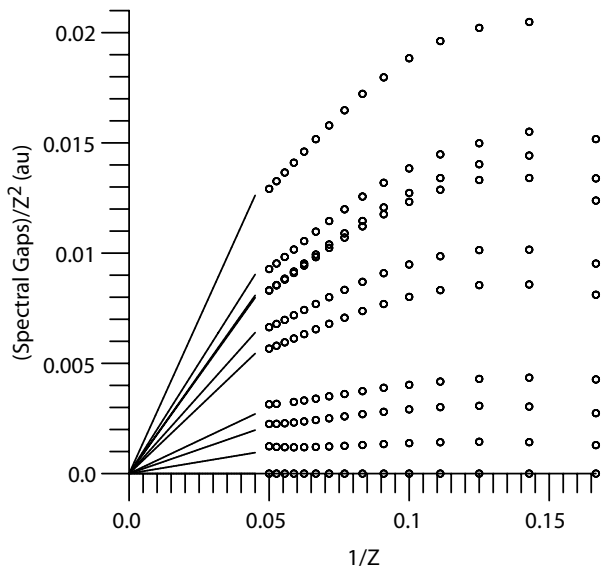}
   }
  \end{center}
  \caption{Splitting of noninteracting Carbon ground state energy by electron interaction. Lines: Asymptotic Schr\"odinger levels (this paper); Circles:
experimental data \cite{NIST,Moore70}. For the highest level at $Z=6$
and the fourth level at $Z=18$, we were unable to find experimental
data.}
  \label{F:Splitting}
\end{figure}

The asymptotic energy levels, despite their excellent orderings,
do not give quantitatively useful spectral gaps for
neutral atoms. However, in the regime of highly positive ions in which they were theoretically
justified (see Theorem \ref{T:isolimit}) they beautifully match the experimental gaps,
as shown in Figure \ref{F:Splitting}.

\noindent
For the Carbon series ($N=6$, $Z=6,7,8,\dots$), we show the experimental spectral gaps
$$
         \frac{E_j(N,Z)}{Z^2} - \frac{E_1(N,Z)}{Z^2}
$$
(circles) and the perturbation-theoretic spectral gaps
$$
         \frac{E_j^{PT}(N,Z)}{Z^2} - \frac{E_1^{PT}(N,Z)}{Z^2} = \frac{\Etilde_j^{(1)}}{Z} - \frac{\Etilde_1^{(1)}}{Z}
$$
(lines) against $\frac{1}{Z}$, with the energy of the lowest level (which shifts with $Z$) having
been subtracted for clarity.
By Theorem \ref{T:isolimit}, the match between PT and Schr\"odinger energy levels would become even
better as $Z$ increases further. But beyond the value $Z=20$ shown here, the match
between Schr\"odinger levels and experiment slowly starts to deviate, due to the onset of relativistic effects, whose
study lies beyond the scope of the present paper.

\subsection{Overall conclusion}
The principal conclusion of this paper is that the semi-empirical hydrogen orbital configurations
of atoms developed by Bohr, Hund and Slater have a precise
mathematical meaning, as asymptotic limits of the true Schr\"odinger ground states for large nuclear charge. (This holds
up to certain small but interesting corrections, as described in Section \ref{Sec:Aufbau}.)
We hope that the limit eigenstates calculated here (see Table \ref{Tab:PTGS})
can serve as a theoretical alternative to semi-empirical discussions of the periodic table
in the literature.

Another use of our findings, as benchmark data for the design and validation of computational methods,
is explored in a companion paper \cite{FG09}.
\\[4mm]
{\bf Acknowledgements} The research of B.G. was supported by a graduate scholarship from EPSRC. We thank
P.Gill for helpful comments, and Ch.Mendl for careful checking of Tables \ref{Tab:PTEigenspaces1},
\ref{Tab:PTEigenspaces2}.


\bibliographystyle{alpha}
\bibliography{PT3}

\newcommand{\etalchar}[1]{$^{#1}$}
\begin{thebibliography}{RJK{\etalchar{+}}07}

\bibitem[AdP01]{Atkins01}
P.~Atkins and J.~de~Paula.
\newblock {\em Atkins' Physical Chemistry, 7th Ed.}
\newblock Oxford University Press, 2001.

\bibitem[AS72]{AbramowitzStegun72}
M.~Abramowitz and I.~A. Stegun.
\newblock {\em Handbook of mathematical functions with formulas, graphs, and
  mathematical tables}.
\newblock National Bureau of Standards Applied Mathematics Series; 55.
  Washington, D.C.: U.S. Dept. of Commerce, 1972.

\bibitem[Boh22]{Bohr22}
N.~Bohr.
\newblock {\em The Theory of Atomic Spectra and Atomic Constitution}.
\newblock Cambridge University Press, 1922.

\bibitem[BS57]{BetheSalpeter57}
H.~A. Bethe and E.~E. Salpeter.
\newblock {\em Quantum mechanics of one-and two-electron atoms}.
\newblock Handbuch der Physik Vol. 35. Springer-Verlag, 1957.

\bibitem[BT86]{Tayloretal1}
Ch.~W. Bauschlicher and P.~R. Taylor.
\newblock Benchmark full configuration-interaction calculations on {H}$_2${O},
  {F}, and {F}$^-$.
\newblock {\em Journal of Chemical Physics}, 85(5):2779--2783, 1986.

\bibitem[Con80]{Condon80}
E.~U. Condon.
\newblock {\em Atomic Structure}.
\newblock Cambridge University Press, 1980.

\bibitem[CS35]{CondonShortley35}
E.~U. Condon and G.~H. Shortley.
\newblock {\em The Theory of Atomic Spectra}.
\newblock Cambridge University Press, 1935.

\bibitem[Dir29]{Dirac29}
P.~A.~M. Dirac.
\newblock Quantum mechanics of many-electron systems.
\newblock {\em Proc. Roy. Soc. London A}, 123:714--733, 1929.

\bibitem[FF77]{FroeseFischer77}
C.~Froese~Fischer.
\newblock {\em The {Hartree-Fock} Method for Atoms. A Numerical Approach.}
\newblock Wiley-Interscience, 1977.

\bibitem[FG09]{FG09}
G.~Friesecke and B.~D. Goddard.
\newblock Asymptotics-based {CI} models for atoms: properties, exact solution
  of a minimal model for {L}i to {N}e, and application to atomic spectra.
\newblock {\em SIAM MMS}, to appear, 2009.

\bibitem[Fri03]{Friesecke03}
G.~Friesecke.
\newblock The multiconfiguration equations for atoms and molecules.
\newblock {\em Archive for Rational Mechanics and Analysis}, 169:35--71, 2003.

\bibitem[FriXX]{Friesecke}
G.~Friesecke.
\newblock {\em In preparation}.
\newblock 20XX.

\bibitem[Gri95]{Griffiths95}
D.~J. Griffiths.
\newblock {\em Introduction to quantum mechanics}.
\newblock Englewood Cliffs, N.J.: Prentice Hall, 1995.

\bibitem[Har28]{Hartree28}
D.~R. Hartree.
\newblock The wave mechanics of an atom with a non-{C}oulomb central field.
  {P}art i -- {T}heory and {M}ethods.
\newblock {\em Proceedings of the Cambridge Philosophical Society}, 24:89--132,
  1928.

\bibitem[Har57]{Hartree57}
D.~R. Hartree.
\newblock {\em The calculation of atomic structures}.
\newblock Wiley, 1957.

\bibitem[Her86]{Herschbach86}
D.~R. Herschbach.
\newblock Dimensional interpolation for two-electron atoms.
\newblock {\em J. Chem. Phys.}, 84(2):838--851, 1986.

\bibitem[Huh93]{Huheey93}
J.~E. Huheey.
\newblock {\em Inorganic chemistry : principles of structure and reactivity}.
\newblock Harper Collins, 1993.

\bibitem[Hun25]{Hund25}
F.~Hund.
\newblock Zur {D}eutung verwickelter {S}pektren, insbesondere der {E}lemente
  {S}candium bis {N}ickel.
\newblock {\em Zeitschrift f\"ur Physik}, 33:345--371, 1925.

\bibitem[Hyl30]{Hylleraas30}
E.~A. Hylleraas.
\newblock {\"U}ber der {G}rundterm der {Z}weielecktronenprobleme von {H$^-$,
  He, Li$^+$, Be$^{++}$} usw.
\newblock {\em Z. Phys.}, 65:209--225, 1930.

\bibitem[Joh05]{NIST05}
R.D. Johnson, editor.
\newblock {\em NIST Computational Chemistry Comparison and Benchmark Database,
  NIST Standard Reference Database Number 101 Release 12}.
\newblock Aug 2005.

\bibitem[Jos02]{Jost02}
J.~Jost.
\newblock {\em Riemannian geometry and geometric analysis}.
\newblock Berlin ; New York : Springer, 2002.

\bibitem[Kat51]{Kat51}
T.~Kato.
\newblock Fundamental properties of {H}amiltonian operators of {S}chr\"odinger
  type.
\newblock {\em Transactions of the AMS}, 70:212--218, 1951.

\bibitem[Kat95]{Kat95}
T.~Kato.
\newblock {\em Perturbation theory of linear operators}.
\newblock Springer-Verlag, 1995.

\bibitem[Lay59]{Layzer59}
D.~Layzer.
\newblock On a screening theory of atomic spectra.
\newblock {\em Ann. Phys. - New York}, 8:271--296, 1959.

\bibitem[Lie84]{Lieb84}
Elliott~H. Lieb.
\newblock Bound on the maximum negative ionization of atoms and molecules.
\newblock {\em Phys. Rev. A}, 29(6):3018--3028, Jun 1984.

\bibitem[LL77]{LanLif}
L.D. Landau and L.M. Lifschitz.
\newblock {\em Quantum mechanics}.
\newblock Pergamon Press, 1977.

\bibitem[Loe86]{Loeser87}
J.~G. Loeser.
\newblock Atomic energies from the large-dimension limit.
\newblock {\em J. Chem. Phys.}, 86(10):5635--5646, 1986.

\bibitem[Moo70]{Moore70}
C.~E. Moore.
\newblock {\em Selected Tables of Atomic Spectra (NSRDS-NBS 3)}.
\newblock Washington, DC: National Bureau of Standards, 1970.

\bibitem[RD71]{RileyDalgarno71}
M.~E. Riley and A.~Dalgarno.
\newblock Perturbation calculation of the helium ground state energy.
\newblock {\em Chem. Phys. Lett.}, 9:382--386, 1971.

\bibitem[RJK{\etalchar{+}}07]{NIST}
Yu. Ralchenko, F.-C. Jou, D.E. Kelleher, A.E. Kramida, A.~Musgrove, J.~Reader,
  W.L. Wiese, and K.~Olsen.
\newblock {\em NIST Atomic Spectra Database (version 3.1.2)}.
\newblock National Institute of Standards and Technology, Gaithersburg, MD,
  2007.

\bibitem[SC62]{SharmaCoulson62}
C.~S. Sharma and C.~A. Coulson.
\newblock Hartree-{F}ock and correlation energies for 1s2s {$^3$S} and {$^1$S}
  states of helium-like ions.
\newblock {\em Proceedings of the Physical Society}, 80:81--96, 1962.

\bibitem[Sch01]{Schwabl01}
F.~Schwabl.
\newblock {\em Quantum Mechanics}.
\newblock Springer, 2001.

\bibitem[Sla30]{Slater30}
J.~C. Slater.
\newblock Atomic shielding constants.
\newblock {\em Physical Review}, 36(1):57--64, 1930.

\bibitem[SO96]{SzaboOstlund96}
A.~Szabo and N.~S. Ostlund.
\newblock {\em Modern Quantum Chemistry}.
\newblock Dover Publications, 1996.

\bibitem[SW67]{SeungWilson67}
S.~Seung and E.~B. Wilson.
\newblock Ground state energy of lithium and three-electron ions by
  perturbation theory.
\newblock {\em J. Chem. Phys.}, 47:5343--5352, 1967.

\bibitem[TRR00]{Tanner00}
G.~Tanner, K.~Richter, and J.~M. Rost.
\newblock The theory of two-electron atoms: between ground state and complete
  fragmentation.
\newblock {\em Rev. Mod. Phys.}, 72(2):497--544, 2000.

\bibitem[TTST94]{Tatewaki94}
H.~Tatewaki, K.~Toshikatsu, Y.~Sakai, and A.~J. Thakkar.
\newblock Numerical {Hartree-Fock} energies of low-lying excited states of
  neutral atoms with {$Z\leq 18$}.
\newblock {\em Journal of Chemical Physics}, 101(6):4945--4948, 1994.

\bibitem[Wil84]{Wilson84}
S.~Wilson.
\newblock Many-body perturbation theory using a bare-nucleus reference
  function: a model study.
\newblock {\em J. Phys. B: At. Mol. Phys.}, 17:505--518, 1984.

\bibitem[Wit80]{Witten80}
E.~Witten.
\newblock Quarks, atoms, and the {1/N} expansion.
\newblock {\em Phys. Today}, 33(7):38, 1980.

\bibitem[Zhi60]{Zhislin60}
G.~M. Zhislin.
\newblock Discussion of the spectrum of {S}chr\"odinger operators for systems
  of many particles.
\newblock {\em Trudy {M}oskovskogo matematiceskogo obscestva}, 9:81--120, 1960.

\end{thebibliography}

\noindent
Address of authors: \\[2mm]
Gero Friesecke \\
Center for Mathematics, TU Munich, Germany, {\tt gf@ma.tum.de} \\[2mm]
Benjamin D. Goddard \\
Mathematics Institute, University of Warwick, U.K. {\tt b.d.goddard@warwick.ac.uk}

\end{document}